\documentclass {aa} 
\usepackage{graphicx}
\usepackage{txfonts}
\usepackage{natbib}
\usepackage{supertabular}
\usepackage{mathrsfs}
\usepackage{fontenc}      
\usepackage{color}  
\usepackage{float} 
  
\begin{document}    

\title{Critical study of the distribution of rotational velocities of Be stars}
\subtitle{II: Differential rotation and some hidden effects interfering with the interpretation of the $V\!\sin i$ parameter}
\author{J. Zorec \inst{1,2}
\and Y. Fr\'emat \inst{1,2,3}
\and A. Domiciano de Souza \inst{4}
\and F. Royer \inst{5}                        
\and L. Cidale \inst{6,7}
\and A.-M. Hubert \inst{5}
\and T. Semaan \inst{8}
\and C. Martayan \inst{9}
\and Y.R. Cochetti \inst{6,7}
\and M.L. Arias \inst{6,7}
\and Y. Aidelman \inst{6,7}  
\and P. Stee \inst{4}    
}
\institute{Sorbonne Universit\'es, UPMC Universit\'e Paris 06 et CNRS UMR 7095, Institut d'Astrophysique de Paris, F-75014 Paris, France
\and 
CNRS UMR 7095, Institut d'Astrophysique de Paris, 98bis Bd. Arago, F-75014 Paris, France; \email{zorec@iap.fr}
\and
Royal Observatory of Belgium, 3 Av. Circulaire, B-1180 Bruxelles, Belgium 
\and
Universit\'e C\^ote d'Azur, Observatoire de la C\^ote d'Azur, CNRS UMR 7293, Lagrange, 28 Avenue Valrose, 06108, Nice Cedex 2, France 
\and
GEPI, Observatoire de Paris, PSL Research University, CNRS UMR 8111, Universit\'e Paris Diderot, Sorbonne Paris Cit\'e, 5 place Jules Janssen, 92190 Meudon, France 
\and
Facultad de Ciencias Astron\'omicas y Geof\'\i sicas, Universidad Nacional de La Plata, Paseo del Bosque S/N, 1900 La Plata, Argentina 
\and 
Instituto de Astrof\'\i sica La Plata, CONICET, 1900 La Plata, 
Argentina
\and
Geneva Observatory, University of Geneva, Maillettes 51, 1290 Sauverny, Switzerland
\and 
European Organization for Astronomical Research in the Southern  Hemisphere, Alonso de Cordova 3107, Vitacura, Santiago de Chile, Chile
}
 
\offprints{J. Zorec: \email{zorec@iap.fr}}    
\date{Received ..., ; Accepted ...,}
\abstract
{}
{We assume that stars may undergo surface differential rotation to study its impact on the interpretation of  {$V\!\sin i$} and on the observed distribution {$\Phi(u)$} of ratios of true rotational velocities {$u=V/V_{\rm c}$} ({$V_{\rm c}$} is the equatorial critical velocity). We discuss some phenomena affecting the formation of spectral lines and their broadening, which can obliterate the information carried by {$V\!\sin i$} concerning the actual stellar rotation.}
{We studied the line broadening produced by several differential rotational laws, but adopted Maunder's expression 
{$\Omega(\theta)=\Omega_o(1+\alpha\cos^2\theta)$} as an attempt to account for all of these laws with the  lowest possible number of free parameters. We studied the effect of the differential rotation parameter $\alpha$ on the measured {$V\!\sin i$} parameter and on the distribution {$\Phi(u)$} of ratios {$u=V/V_{\rm c}$}.}
{We conclude that the inferred {$V\!\sin i$} is smaller than implied by the actual equatorial linear rotation velocity {$V_{\rm eq}$} if the stars rotate with {$\alpha<0$}, but is larger if the stars have {$\alpha>0$}. For a given {$|\alpha|$} the deviations of {$V\!\sin i$} are larger when {$\alpha<0$}. If the studied Be stars have on average {$\alpha<0$}, the number of rotators with {$V_{\rm eq}\simeq0.9V_{\rm c}$} is larger than expected from the observed distribution {$\Phi(u)$}; if these stars
have on average {$\alpha>0$}, this number is lower than expected. We discuss seven phenomena that contribute either to narrow or broaden spectral lines, which blur the information on the rotation carried by  {$V\!\sin i$} and, in particular, to decide whether the Be phenomenon mostly rely on the critical rotation. We show that two-dimensional radiation transfer calculations are needed in rapid rotators to diagnose the stellar rotation more reliably.} 
{}
\titlerunning{Rotation of Be stars}
\authorrunning{J. Zorec et al.} 

\maketitle 

\section{Introduction}\label{intro}   

  In Paper I of this series \citep{zor16a} we have obtained the distribution $\Phi(V/V_{\rm c})$ of true velocity ratios $V/V_{\rm c}$ ($V_{\rm c}$ is the critical linear velocity at the equator) corresponding to a sample of 233 galactic classical Be stars. This distribution was derived from that of apparent parameters $V\!\sin i$,
which were corrected for measurement uncertainties, and assuming that the inclination angles $i$ were distributed at random. Since Be stars are rapid rotators, the parameters $V\!\sin i$ were corrected for the Stoeckley underestimation (see definition in Paper I) in the frame of rigid rotation using the original von Zeipel (1924) theorem. In this way we ensure that the gravity darkening (GD) effect is not underestimated. Our aim was to determine an upper limit of rotational velocities of Be stars by taking some effects that can disturb the determination of the true velocity $V$ into account. By overestimating the Stoeckley corrections on purpose, the inferred $\Phi(V/V_{\rm c})$ distribution has the largest possible mode, i.e., the distribution reveals the greatest possible velocity ratios $V/V_{\rm c}$ that might characterize the Be phenomenon. Then, we corrected the distribution of $V\!\sin i/V_{\rm c}$ ratios for overestimations of the $V\!\sin i$ values induced by macroturbulent motions in the stellar atmospheres and distribution of ratios $V/V_{\rm c}$ \footnote{The correction for a presumable effect associated with the orbital motion of Be stars in binary systems presented in Paper I must not be taken into account (see Appendix~\ref{nobin}).}.\par
  In Paper I we have assumed implicitly that the angular velocity $\Omega$ on the surface of stars is uniform. In the present approach, we take the surface angular velocity dependent on the colatitude $\theta$, so that the measured $V\!\sin i$ depends on the characteristics of the surface rotation law $\Omega=\Omega(\theta)$.\par 
  The first part of the present contribution (Sect.~\ref{dif_rot}) aims at exploring two questions: What is the effect of the surface differential rotation on the measured $V\!\sin i$ parameters? and What could be the deviation
of the observed distribution of the true velocities ratios $V/V_{\rm c}$ from the distribution of actual equatorial rotational velocities $V_{\rm eq}/V_{\rm c}$? In the second part of this work (Sect.~\ref{uncert_vsini}) we discuss some measuring and conceptual uncertainties that affect the determination of $V\!\sin i$. These uncertainties can cast doubts on the quality of the information that this parameter may carry on the actual properties of the stellar surface rotation.\par   

\section{Effect carried by the surface differential rotation} \label{dif_rot} 
\subsection{Is differential rotation a timely topic for Be stars?}\label{why_difrot}

  Independent of the internal rotation law of massive stars, either conservative or non-conservative ``shellular", \citet{clem79} and \citet{mae08} have shown that the two convection zones in the envelope associated with increased opacity due to He and Fe ionization can be considerably enlarged in depth from rapid rotation. These regions together establish an entire convective zone beneath the surface that spread out over a non-negligible region: from 1/8 of the stellar radius at the pole to nearly 1/4 at the equator \citep{mae08}. Since the rotation law in these layers can have a direct incidence on the surface rotation, we are interested in its nature. In general, two extreme approximations have currently been used to account for the angular momentum distribution in convective regions: first, redistribution promoted by the turbulent viscosity \citep{mae08}, which ends up establishing rigid rotation; and, second, redistribution of specific angular momentum $j\!=\!\Omega\varpi^2$ carried by the convective plumes \citep{tay73}, which favors the existence of rotation laws characterized by $j\simeq$ constant \citep{deup98,deup01}.  The solar convective regions, however, which are not only characterized by significant turbulence but also rotate differentially \citep{sch98}, do not conform with any intermediate frame between these two extreme possibilities.\par
\begin{figure*}[ht!]
\centerline{\includegraphics{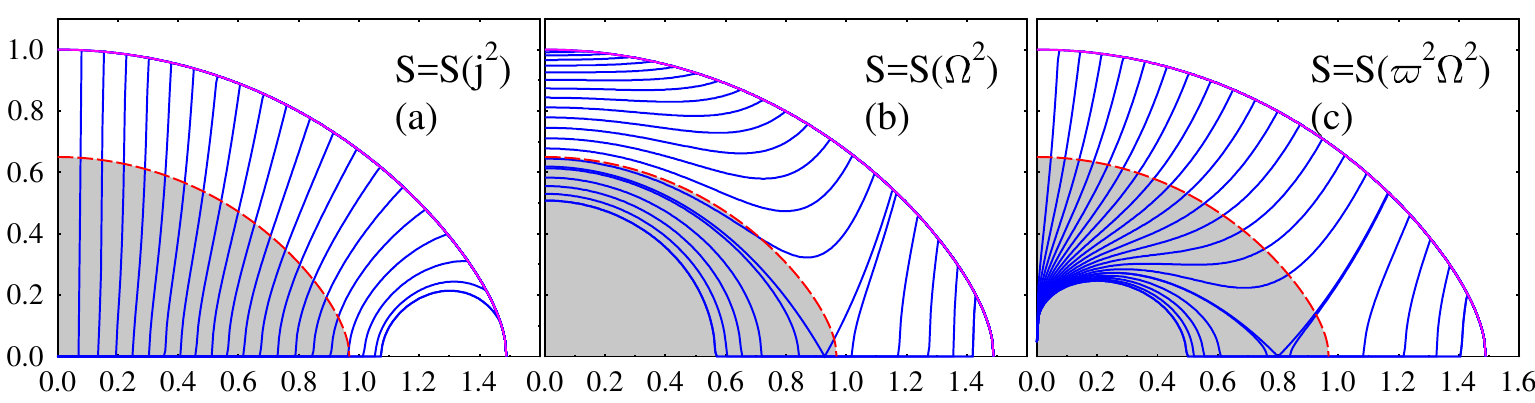}}
\caption{\label{fig_13} Curves of internal constant angular velocity $\Omega(r,\theta)\!=\!const$ in the stellar envelope (blue lines). a) Curves corresponding to the condition $S\!=\!S(j^2)$; b) curves obtained with $S\!=\!S(\Omega^2)$; and c) curves calculated for $S\!=\!S(\varpi^2\Omega^2)$. The solutions are valid in the convective envelope, i.e., above the shaded central region [$S$ = specific entropy; $j$ = specific angular momentum; and $\varpi$ = distance to the rotation axis]. Ordinates coincide with the rotation axis; abscissas are in the equatorial plane.}
\end{figure*}

 
  According to the solar rotational picture, where the layers are unstable to convection and rotate differentially with a non-conservative non-shellular pattern, we may wonder whether  some coupling can also exist between convection and rotation beneath the surface in rapidly rotating massive and intermediate
mass stars. Solving the baroclinic balance relation obtained with the curl of the time-independent momentum equation of an inviscid, axisymmetric rotating star without magnetic fields, \citet{zor11} obtained solutions for the angular velocity distribution in the envelope under several conditions: 1) the surfaces of specific entropy $S$ are parallel to the surfaces of specific angular momentum $j$, $S\!\!\!=\!S(j^2)$; 2) the surfaces of specific entropy and angular velocity coincide, i.e., $S\!=\!S(\Omega^2)$; and 3) the surfaces of specific entropy $S$ are parallel with the surfaces of constant specific rotational kinetic energy, $S\!=\!S(\varpi^2\Omega^2)$. In Fig.~\ref{fig_13} we show some examples of curves $\Omega(r,\theta)\!=\!const$ that obey these assumptions. Should this phenomenon take place, its imprint on the outermost stellar layers would certainly translate into an angular velocity dependent on the latitude. The existence of a possible external differential rotation in Be and other massive and intermediate mass stars was speculated so far by a number of authors \citep[cf.][]{sto68b,zor86,zor87,sto87,zo90nato316,cra93,rieu07,zor12,
zor16b}. \par
  From all of the above, the interest of studying the external angular velocity law resides in the fact that it may carry information on the sub-photospheric stellar structure. However, as the characterization of a star also needs to account for its $V\!\sin i$ parameter,  differential rotation affects  the Doppler broadening of spectral lines in a particular way, and this has to be taken into account. In Sect.~\ref{rot_broad} we show that depending on the characteristics of the external rotation law, the larger contribution to the Doppler broadening can be produced anywhere in the observed stellar hemisphere, and not necessarily in the equator. Finally, rotation laws similar to those depicted in Fig.~\ref{fig_13} not only determine the stellar external geometry, but can also induce gravity darkening laws that strongly deviate from the classical von Zeipel description (see Sect.~\ref{comm_gd}).\par

\subsection{Rotational line broadening}\label{rot_broad}

  We adopt a Cartesian reference system $(x,y,z)$ centered on the star, where the $z-$axis is directed positively toward the observer and the $x-$ and $y-$axis are in the plane of the sky. Assuming that the rotation axis is in the $yz-$plane, the loci of points contributing to a same Doppler displacement $\Delta\lambda$ is described by the $z$ component of the vector $\vec{\Omega}\wedge\vec{R,}$

\begin{equation}
\Delta\lambda = (\lambda/c)(\vec{\Omega}\wedge\vec{R})_z = (\lambda/c)\Omega(\theta)x\sin i,
\label{eq21}
\end{equation}

\noindent where $\Omega(\theta)$ is the angular velocity on the stellar surface, $\theta$ is the colatitude and $\vec{R}$ is the position vector, and $i$ is the inclination angle measured from the stellar polar axis and the direction toward the observer. To characterize the Doppler displacements described by Eq.~(\ref{eq21}) more clearly, we normalize this relation by the displacement produced in the limb at the equator,

\begin{equation}
\mathscr{C} = \frac{\Delta\lambda}{\Delta\lambda_{\rm e}} = \left[\frac{\Omega(\theta)}
{\Omega_{\rm e}}\right]\frac{x}{R_{\rm e}},
\label{eq22}
\end{equation}

\noindent where $\Omega_{\rm e}$ is the angular velocity at the equator and $\mathscr{C}$ is a constant independent of $\sin i$ that represents an isoradial velocity curve. Curves $\mathscr{C}=\pm|\mathscr{C}|$ are symmetric with respect to the $yz$-plane. Considering only the $x\geq0$ side, we have $0\leq\mathscr{C}\leq\mathscr{C}_{\rm max}$.  The $\mathscr{C}=0$ curve is on the plane containing the $z-$direction and the rotation axis whatever the rotation law.  The value $\mathscr{C}=1$  corresponds to the stellar limb in the equator (i.e., $x_{\rm max}=R_{\rm e}$) either for a rigid rotation or for $\Omega(\theta)$ accelerated from the pole toward the equator. Finally, when $\Omega(\theta)$ is accelerated from the equator toward the pole, it happens that $\mathscr{C_{\rm max}}=\mathscr{C}(x_{\rm max})\geq1$ with $x_{\rm max}\leq1$.\par  

\begin{figure*}[ht!]
\centerline{\includegraphics{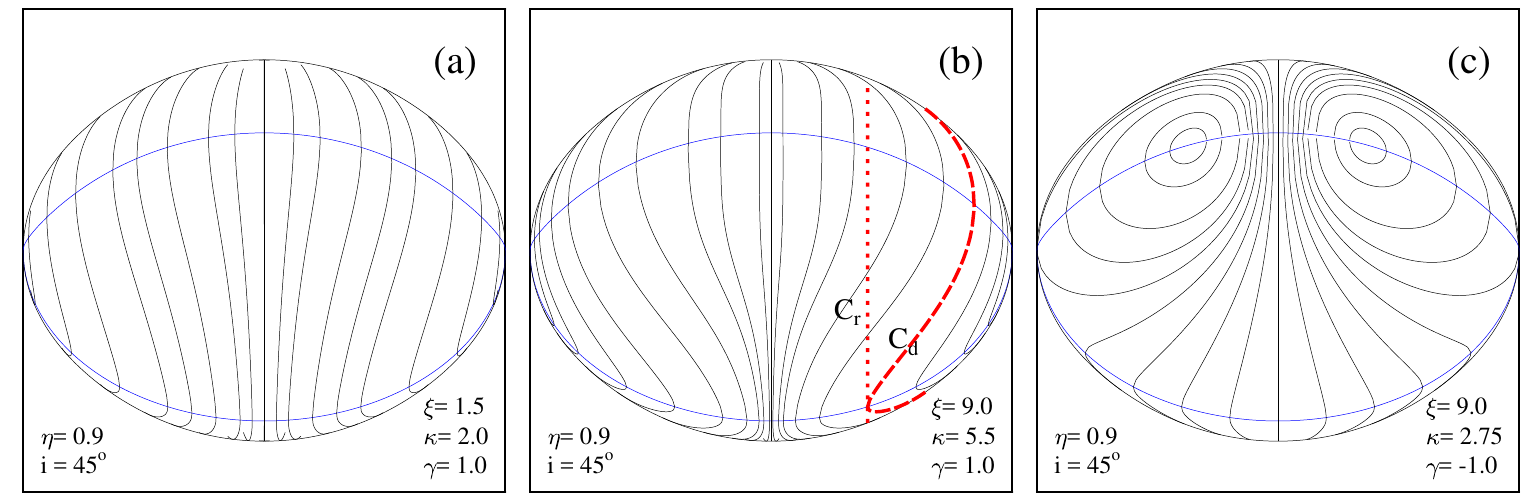}}
\caption{\label{fig_14} Curves of constant radial velocity $\mathscr{C}$ contributing to the rotational Doppler broadening of spectral lines. a) $\mathscr{C}-$curves for a Maunder-like surface angular velocity law with $\partial\Omega(\theta)/\partial\theta>0$; b) $\mathscr{C}-$curves for a law given by Eq.~(\ref{eq23}) with $\gamma>0$, so that $\partial\Omega(\theta)/\partial\theta>0$; c) $\mathscr{C}-$curves for a law given by Eq.~(\ref{eq23}) with $\gamma<0$, so that $\partial\Omega(\theta)/\partial\theta<0$. In cases a) and b) the largest Doppler displacement is produced in the limb at the equator, while in case c) is produced in the center of the upper closed curves. Case b) identifies the lines producing the same Doppler displacement, $\mathscr{C}_{\rm r}=\mathscr{C}_{\rm d}$, where $\mathscr{C}_{\rm r}$ (red dotted line) indicates rigid surface rotation, and $\mathscr{C}_{\rm d}$ (red dashed line)  a differential surface rotation. In all cases the parent Doppler displacement from rigid rotation is determined by the straight line touching the corresponding $\mathscr{C}_{\rm d}$ curve at the equator.}
\centerline{\includegraphics{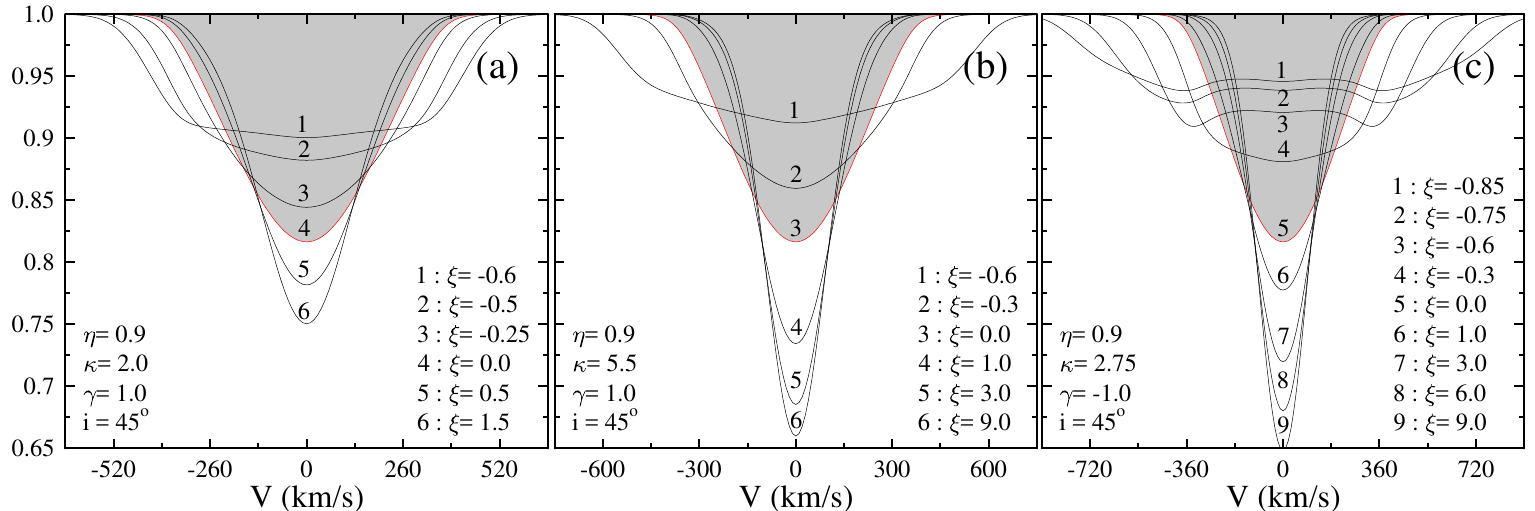}} 
\caption{\label{fig_15} Rotationally broadened line profiles of an isolated Gaussian line, whose intensity and equivalent width depends on the effective temperature and gravity such as a \ion{He}{i}\,4471 transition. Calculations include gravitational darkening. The parent non-rotating object has $T_{\rm eff}=22000$ K and $\log g=$ 4.0. The angular velocity law is given by Eq.~(\ref{eq23}) with the same parameters ($\kappa,\gamma,\eta,i$) as in the corresponding block in Fig.~\ref{fig_14}, except that $\xi$ has a wider range of values. The case $\xi=0$ represents the rigid rotation and the corresponding line profile is highlighted with shading. In all blocks the ordinates are the same.}
\end{figure*}

  To discuss more general rotation laws, let us assume the following expression

\begin{equation}
\Omega(\theta) = \Omega_{\rm p}(1+\xi\varpi^{\kappa})^{\gamma},
\label{eq23} 
\end{equation} 

\noindent where $\varpi=r(\theta)\sin\theta$ is the distance to the rotation axis; $r(\theta)=R(\theta)/R_{\rm e}$ is the normalized radius-vector that describes the stellar surface, $\Omega_{\rm p}$ is the polar angular velocity, and $\Omega_{\rm e}=\Omega_{\rm p}(1+\xi)^{\gamma}$ is the equatorial angular velocity; $\xi$, $\kappa$ and $\gamma$ are constants. We can readily see that $\partial\Omega/\partial\theta\gtrless0$ when  $\xi\times\gamma\times\kappa\gtrless0$, respectively. When $\partial\Omega/\partial\theta<0,$ it is $x_{\rm max}=[-1/\xi(1+\kappa\gamma)]^{1/\kappa}$, which leads to $\mathscr{C_{\rm max}}=\mathscr{C}(x_{\rm max})\geq1$ on the meridian that contains the $x-$axis. We also have $\Omega_{\rm max}=\Omega(x_{\rm max})=\Omega_{\rm p}[\gamma\kappa/(1+\gamma\kappa)]^{\gamma}\geq\Omega_{\rm e}$, which for appropriate values of the constants $(\xi,\gamma,\kappa)$ can imply $\Omega_{\rm max}x_{\rm max}\geq\Omega_{\rm e}R_{\rm e}$ in the first meridian somewhere above the equator. The condition $\Omega_{\rm max}x_{\rm max}\leq\Omega_{\rm c}(x_{\rm max})x_{\rm max}$, where $\Omega_{\rm c}(x_{\rm max})$ is the local critical angular velocity, imposes constraints on the values of the constants $(\alpha,\gamma,\kappa)$. \par 
   Adopting $\kappa=2$ and $\gamma=1$, Eq.~(\ref{eq23}) becomes \par\begin{equation}
\Omega(\theta)=\Omega_1(1-\epsilon\cos^2\theta),
\label{eq24}
\end{equation}
 
\noindent where $\Omega_1=\Omega_{\rm p}[1+\xi r^2(\theta)]$ and $\epsilon=\xi r^2(\theta)/[1+\xi r^2(\theta)]$. For slowly rotating stars $r(\theta)\simeq1$, $\epsilon$ and $\Omega_1$ can be considered constants, so that the relation in Eq.~(\ref{eq24}) becomes the known Maunder law.\par
  When the angular velocity in the surface is uniform [$\Omega(\theta)\!=$ constant $\forall~\theta$], from Eq.~(\ref{eq22}) we find that the isoradial velocity curves, i.e., $\mathscr{C}=$ constant, are straight lines parallel to the $y-$axis. Otherwise, for $\Omega=\Omega(\theta)$ the isoradial velocity curves $\mathscr{C}$ resemble those shown in Fig.~\ref{fig_14}. Figures~\ref{fig_14}a and \ref{fig_14}b show $\mathscr{C}=$ constant for angular velocities accelerated from the pole toward the equator, while Fig.~\ref{fig_14}c shows a case where $\Omega(\theta)$ is accelerated from the equator toward the pole. In cases ``a" and ``b", the maximum Doppler displacement is produced at the stellar limb in the equator, while in case ``c" this maximum is produced in the middle of the closed circles 
(`owl' eyes) on the visible part of the meridian contained in the $(x\,\hat{\iota},\vec{\Omega})-$plane. In this case, the curve $\mathscr{C}=1$ is also closed and contains the equatorial limb point, while all the remaining curves for $\mathscr{C}>1$ are closed.\par
  Figure~\ref{fig_14}b shows two curves (red dotted and dashed curves) that produce the same Doppler displacement, i.e. ,$\mathscr{C}_{\rm r}=\mathscr{C}_{\rm d}$, where $\mathscr{C}_{\rm r}$ denotes the radial velocity for rigid rotation (red dotted  line) and $\mathscr{C}_{\rm d}$ is for differential rotation (red dashed curve). By definition, both isoradial velocity curves coincide at the equator, nevertheless, the $\mathscr{C}_{\rm d}$ curve is longer than for $\mathscr{C}_{\rm r}$. This is the main source of difference detected in residual intensities of absorption lines broadened by rigid and differential rotation laws.\par 
  Figure~\ref{fig_15} shows rotationally broadened line profiles of a hypothetical Gaussian line, whose intensity and equivalent width depend on the effective temperature and gravity as the actual \ion{He}{i}\,4471 line. In all depicted cases, the GD was taken into account. The parent non-rotating object corresponds to $T_{\rm eff}=22000$ K and $\log g=$ 4.0. The angular velocity laws used for these lines are the same as those in Fig.~\ref{fig_14}, except for the parameter $\xi$, which ranges from negative to positive values given in the inlaid in the lower right corners of Fig.~\ref{fig_15}. The cases with $\xi=0$ are for rigid rotation and the corresponding line profiles are highlighted with shading to ease the comparison. Lines calculated with $\xi\leq0$ have FWHM (full width at half maximum) wider than FWHM$_{\rm rigid}$, while for $\xi>0$ the FWHM are smaller than FWHM$_{\rm rigid}$.\par
  In Fig.~\ref{fig_15}c, the line profiles with $\xi<0$  have a central emission-like component. This phenomenon is produced by the isoradial velocity $\mathscr{C}_{\rm d}$ curves close to $\mathscr{C}_{\rm d}=0$, which are shorter than those approaching $\mathscr{C}_{\rm d}=1$. The absorption in the central wavelengths is then less pronounced and the line profile acquires an emission-like aspect. In case ``c", when the gradient $\partial\Omega/\partial\theta<0$ is large enough, the polar regions become dimpled. All calculations in Figs.~\ref{fig_14} and \ref{fig_15} were performed for the equatorial centrifugal to gravitational force ratio $\eta=0.9$, where $\eta$ is
 
\begin{equation}
\eta=u^2[R_{\rm e}/R_{\rm c}]
\label{eq29-1}
\end{equation}
  
\noindent and $R_{\rm e}$ and $R_{\rm c}$ are the stellar equatorial and critical radii, respectively.  Frequently, we use the parameter $\eta$ instead of $\Omega_{\rm p}$ or $\Omega_{\rm e}$ to quantify the rotation rate, however, Eq.~(\ref{eq23}) can also be written as 

\begin{equation}
\frac{\Omega(\theta)}{\Omega_{\rm c}}=\left[\frac{\eta}{R_{\rm e}(\eta)/R_{\rm c}}\right]^{1/2}\left(\frac{1+\xi\varpi^k}{1+\xi} \right)^{\gamma}  
\label{eq29-n}
,\end{equation}

\noindent where $R_{\rm e}(\eta)/R_{\rm c}$ is calculated as explained in \citet{zor11} and \citet{zor12}. We note that  $R_{\rm e}=R_{\rm e}(\eta,\xi,k,\gamma,M,t)$, where $M$ is the stellar mass and $t$ its age. To calculate the geometrical
shape of rotating stars and the line profiles in Figs.~\ref{fig_14} and \ref{fig_15}, respectively, we used $M/M_{\odot}=8.2$ and $t/t_{\rm MS}=0.72$ ($t_{\rm MS}$ is the time spent by a star in the main-sequence evolutionary phase), which compared to a parent non-rotating object corresponds to $T_{\rm eff}= 22\,000$~k and $\log g=4.0$.\par
  In what follows, we illustrate the effect of the differential rotation on the determination of the  $V\!\sin i$ parameter
and its incidence on the distribution of the true equatorial rotational velocity.\par 

\begin{figure}[]
\centerline{\includegraphics[scale=0.85]{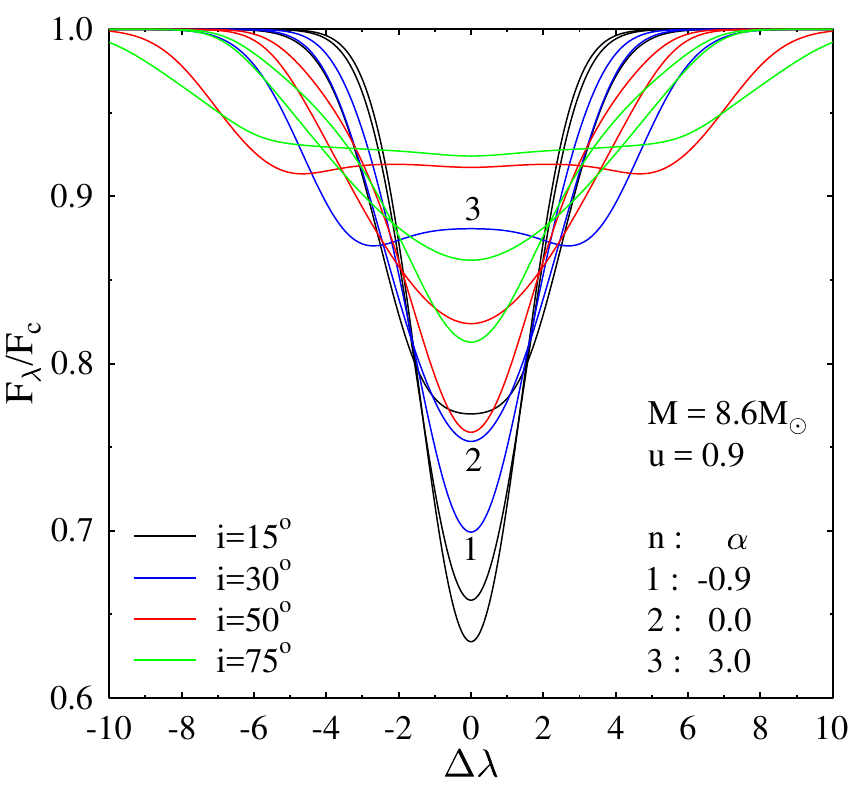}}
\caption{\label{fig_15u} Rotationally broadened line profiles of an intrinsic Gaussian absorption line, whose intensity and equivalent width depends on the effective temperature and gravity like the \ion{He}{i}\,4471 transition. Calculations include gravitational darkening in an object rotating at the equator with $u\!=\!V/V_{\rm c}\!=\!0.9$, whose parent non-rotating counterpart has $M=8.6M_{\odot}$ and $t/t_{\rm MS}=0.5$. The angular velocity law is given by Eq.~(\ref{eq27}). The colors indicate the inclination angles, but in this figure only the $i\!=\!30\degr$ (blue) corresponding to different values of $\alpha$ are explicitly identified.}
\end{figure}

\subsection{Effect on the $V\!\sin i$ parameter}\label{eff_vsini}

   In Sect.~\ref{why_difrot} we pointed out that owing to our ignorance about the physical circumstances that may produce differential rotation in stars, no specific rotational law can be preferred to calculate the spectral line broadening. The first reasonable step we can undertake to uncover  the trends produced by possible components of differential rotation is to adopt a $\Omega(\theta)$ law with the fewest possible number of free parameters. We have therefore adopted the known Maunder relation, 

\begin{equation}
\Omega(\theta,\alpha) = \Omega_{\rm e}(1+\alpha\cos^2\theta),
\label{eq27}
\end{equation} 

\noindent which comes from the generic form given by Eq.~(\ref{eq23}) by putting $k=2$ and $\gamma=1$, and considering that in Eq.~(\ref{eq24}) $\Omega_1$ and $\epsilon$ are constants, so that $\Omega_{\rm e}=\Omega_1$, $\alpha=-\epsilon$ and \par 

\begin{equation}
\Omega_{\rm e} = \Omega_{\rm c}\eta^{1/2}[R_{\rm c}/R_{\rm e}(\eta)]^{3/2}.
\label{eq27bis}
\end{equation}

   In Eq.~(\ref{eq27}), $\alpha$ is the free {\sl differential rotation parameter}, whose possible values are limited 
in this work to the interval $-1<\alpha<+\infty$. In principle, we could also impose values $\alpha<-1$, but in our initial research on the effects produced on the broadening of lines, we avoid the case where the pole and equator rotate in opposite senses by imposing the condition $\alpha>-1$. On the other hand, because the distance to the rotation axis diminishes as $\theta\to0$, there is no limitation to the positive values of $\alpha$. Obviously, rigid rotation is described by $\alpha=0$. From Eq.~(\ref{eq27}) it follows that $\partial\Omega/\partial\theta>0$ if $\alpha<0$, which produces $\delta<0$, and it is $\partial\Omega/\partial\theta<0$ when $\alpha>0$, which implies $\delta>0$. Since $0\leq v\leq1$ and $0\leq u\leq1$, from Eq.~(\ref{eq26}) it follows that $-1\leq\delta\leq1$.\par
   Almost the entire variety of line profiles calculated with Eq.~(\ref{eq23}) can reasonably be accounted for with the Maunder relation [Eq.~(\ref{eq27})], as shown in Fig.~\ref{fig_15u}. For these profiles we used an intrinsic Gaussian absorption line assimilated to the \ion{He}{i}\,4471 transition in an object with $M=8.6M_{\odot}$ and $t/t_{\rm MS}=0.5$. The line broadening was calculated for the equatorial velocity ratio $V/V_{\rm c}=0.9$ and for several values of $\alpha$ and inclination angles.\par          
  Two methods are currently used to determine the $V\!\sin i$ parameter:\par
\medskip  
1. {\sl FWHM against $V\!\sin i$}\par
\medskip
 This simple method is based on a polynomial relation between the FWHM  of spectral lines that are not strongly affected by the Stark broadening, with the $V\!\sin i$ determined for well-studied stars or with synthetic line profiles of rigidly rotating model atmospheres \citep{struve30,slt55,slett75,slett82}. \par
\begin{figure*}[]
\centerline{\includegraphics[scale=1.1]{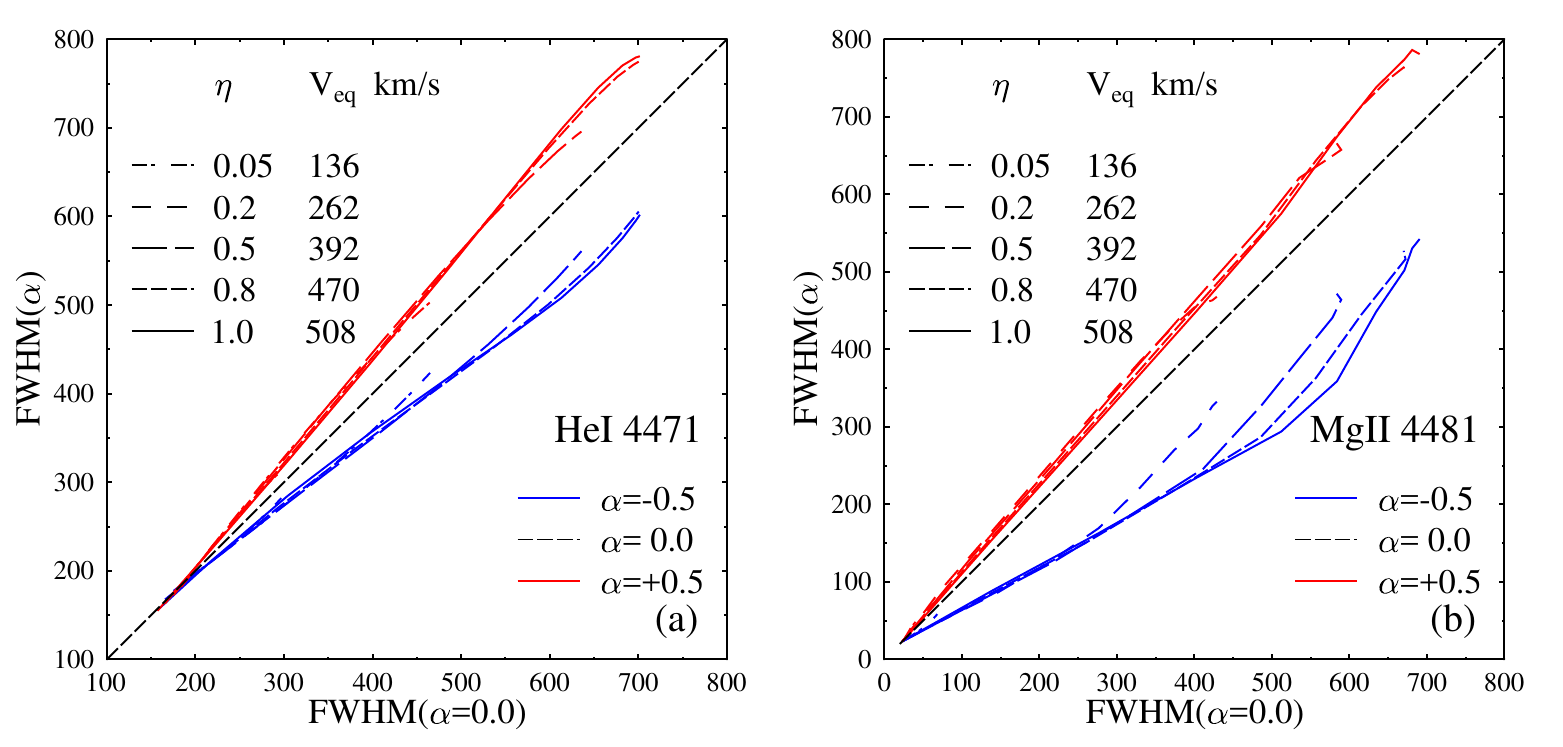}} 
\caption{\label{fwhm_p} a) FWHM  in km~s$^{-1}$ observed at different angles $i$ of the HeI\,4471 line broadened by differential rotation characterized by different values of the parameter $\alpha$ in an object of "pnrc" parameters $T_{\rm eff}=23\,000$ and $\log g=4.1$ rotating at different rates $\eta$. In abscissas are the FWHM of the same line corresponding to rigid rotators having the same rates $\eta$. b) Idem for the MgII\,4481 line.}
\end{figure*}

  Taking the \ion{He}{i}\,4471 and \ion{Mg}{ii}\,4481 lines as witnesses of different sensitivities to physical conditions of line formation in stellar atmospheres (see also discussion in Sect.~\ref{rad_sf}), we calculated their theoretical FWHM broadened using $\Omega(\theta)$ given in Eq.~\ref{eq27} for several values of $\alpha$ including $\alpha=0$. The results are shown in Fig.~\ref{fwhm_p}, where the sensitivity of FWHM to the differential parameter $\alpha$ is apparent. This clearly shows that if differential rotation exists, depending on the sign of $\alpha$, the $V\!\sin i$ parameters determined using the above FWHM-based method barely reflects the true equatorial rotation of stars. As reference to inquire as to what could be the expected effects in stars, it is worth noting that for the Sun it is $\alpha=-0.3$.\par 
  
\medskip
2. {\sl Fourier transform}\par 
\medskip
 The rather widespread method based on the Fourier transform (FT) to determine the $V\!\sin i$ assumes that an observed line profile $\mathscr{F}(\lambda)$ can be represented by the convolution of the flux line profile $F(\lambda)$ of a non-rotating star with a `rotation broadening function' $G_{\rm R}(\lambda)$. This method requires that the specific intensity $I(\lambda,\mu)$ contributing to the observed line flux $F(\lambda)$ has the same shape over the entire stellar disc, which is possible only if the temperature and gravity are uniform over the entire stellar disc. The analytical expression of $G_{\rm R}(\lambda)$, which is independent of the inclination angle, also requires that the star be spherical and behave as a rigid rotator \citep{gray1,gray2}. The $V\!\sin i$ is finally obtained by comparing the zeroes of the FT of flux line profile $\mathscr{F}(\lambda)$ with the corresponding zeroes of the FT of the function $G_{\rm R}(\lambda)$. \par 
 However, the surface of a rotating star is neither spherical nor does it have uniform effective temperature and gravity. From Fig.~\ref{fig_14} it is clearly seen that in a differential rotator the lines of constant radial velocity contributing to a given rotational Doppler displacement impose a rather complicated analytical representation, which is dependent on the a priori unknown inclination angle-$i$ \citep{huang_1961}. We call $G_{\rm D}(\lambda)$ a hypothetical ``rotation broadening function" that we can derive for a given surface rotation law. Even if we wanted to maintain a specific intensity $I(\lambda,\mu)$ behaving as in a non-rotating star, the zeroes of the FT of the function $G_{\rm D}(\lambda)$ would not correspond to those derived from the FT of $G_{\rm R}(\lambda)$ currently used in the literature. So, the zeroes of the FT of $\mathscr{F}(\lambda)$ are not compatible with those of FT[$G(\lambda)$]. As a consequence, the $V\!\sin i$ that is derived using the classical FT method cannot be considered a reliable parameter to represent the rotation of a given star.\par
\begin{figure*}[]
\centerline{\includegraphics[scale=1.0]{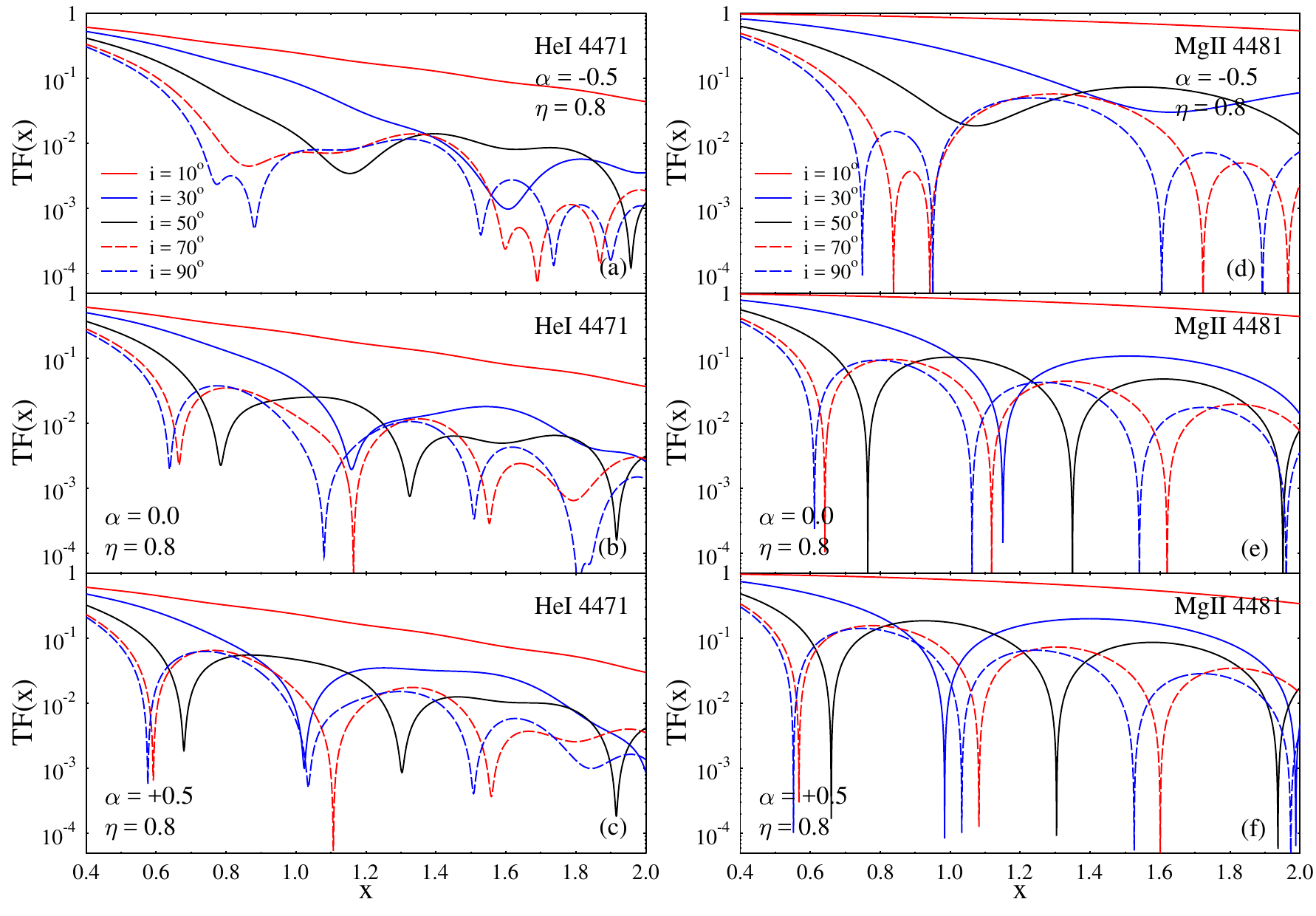}}
\caption{\label{tf_p} a), b), c) Fourier transforms (FT) of the HeI\,4471 line broadened by differential rotation characterized by $\alpha=-0.5$, 0.0 and +0.5, respectively, in an object of "pnrc" parameters $T_{\rm eff}=23\,000$ and $\log g=4.1$ rotating at $\eta=0.8$. d), e), f) Idem for the MgII\,4481 line.}
\end{figure*}

\begin{figure*}[]
\centerline{\includegraphics[scale=1.1]{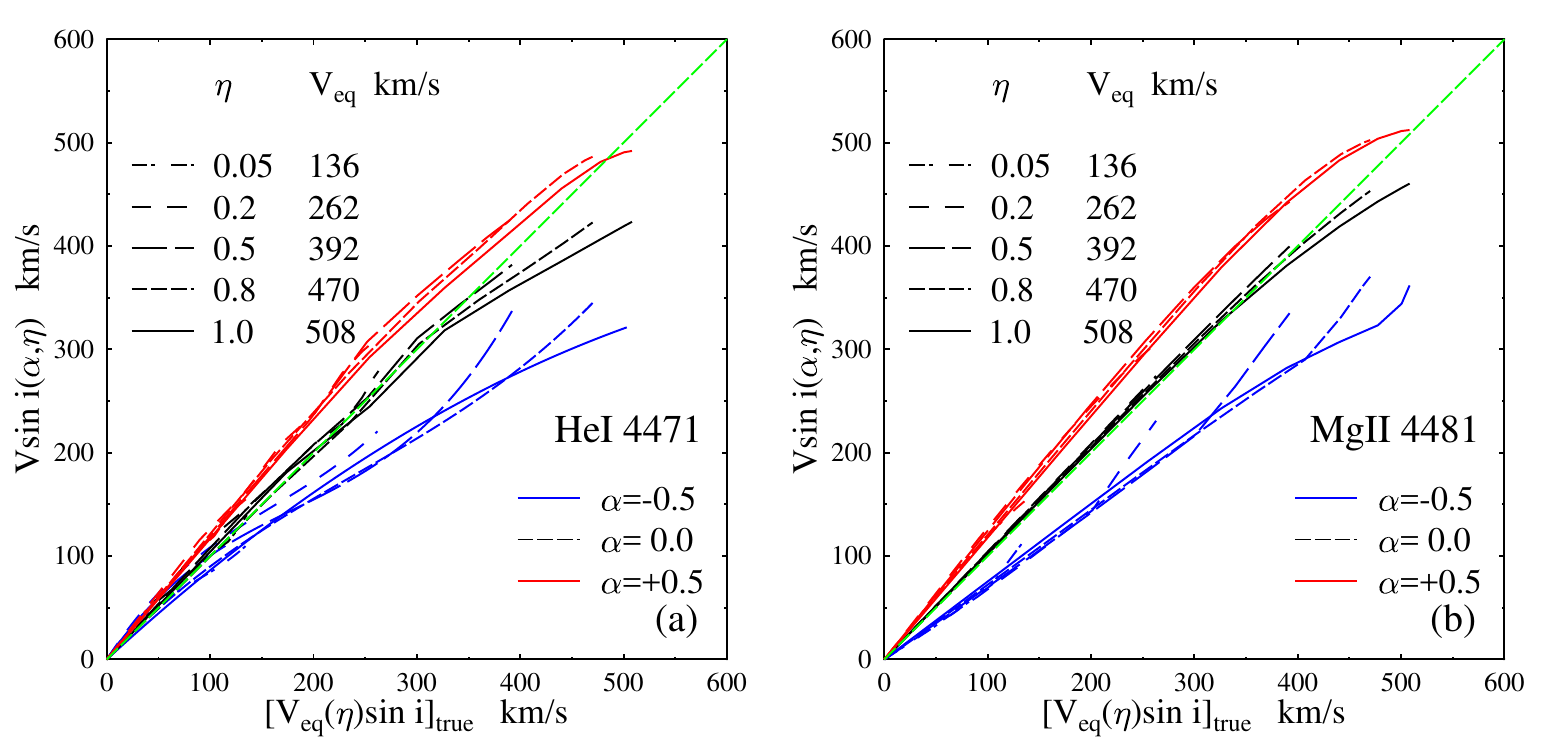}}
\caption{\label{vsini_p} a) Measured $V\!\sin i$ parameters in km~s$^{-1}$ by means of the FT technique of the HeI\,4471 line broadened by differential rotation characterized by different values of the parameter $\alpha$ in an object of "pnrc" parameters $T_{\rm eff}=23\,000$ and $\log g=4.1$ rotating at different rates $\eta$. In abscissas are the true equatorial $V\!\sin i$ for the same rates $\eta$. b) Idem for the MgII\,4481 line. }
\end{figure*}

 
  We calculated \ion{He}{i}\,4471 and \ion{Mg}{ii}\,4481 lines broadened by differential rotators and calculated 
their FT. Depending on the spectral line and the value of $\alpha$, the FT of these lines may have unusual shapes where the zeroes are difficult to interpret. In Fig.~\ref{tf_p} are shown some FT of the \ion{He}{i}\,4471 and \ion{Mg}{ii}\,4481 lines broadened by differential rotators characterized by the pnrc parameters $T_{\rm eff}=$ 23\,000 K, $\log g=$ 4.1, different values of $\alpha$, where all have the same equatorial velocity rate $\eta=0.8$. The resulting $V\!\sin i$ parameters obtained using the classical FT method are compared in Fig.~\ref{vsini_p} with the actual $V_{\rm eq}\!\sin i$ corresponding to several values of $\eta$, parameters $\alpha$, and different inclination angles. Deviations to the $y=x$ also exist for rigid rotators. These deviations are due to the gravitational darkening effect, which in the classical rigid-rotation broadening function $G_{\rm R}(\lambda)$ is not taken into account. If differential rotation actually existed and if it were at least on the order of that observed in the Sun ($\alpha\simeq-0.3$), the $V\!\sin i$ obtained with the FT method would largely deviate from those which are currently believed to characterize the stellar equatorial rotation. \par 

  Figures~\ref{fwhm_p} and \ref{vsini_p} suggest that the $V\!\sin i$ obtained using either the FWHM or the FT method can be overestimated when $\alpha>0$, or underestimated for $\alpha<0$ as compared with its value for rigid rotators.  Also, for a given $|\alpha|$ value the deviation is larger when $\alpha<0$. This means that if stars were actually differential rotators, the distributions derived thus far of ratios $V/V_{\rm c}$ would obviously not represent the distributions of actual equatorial rotations. In the following section we explore the effects induced by the differential rotation on the the distribution of $V/V_{\rm c}$ ratios .\par   

\subsection{Effect on the distribution of ratios $V/V_{\rm c}$}\label{eff_phiu} 
\subsubsection{Some preliminary relations}\label{prelim_cal} 

   We assume that a given studied star undergoes differential rotation and that $V_{\rm D}\!\sin i$ is its projected rotational parameter derived by one of the methods evoked above. This quantity can be written as

\begin{equation}
V_{\rm D}\!\sin i = V_{\rm e}\!\sin i + \Delta V(\sin i),
\label{eq25}
\end{equation}
 
\noindent where $V_{\rm e}$ is the equatorial velocity of the star assumed to rotate as a rigid rotator, and $\Delta V(\sin i)$ is the contribution to the spectral line broadening by the hemisphere-averaged deviation from rigid rotation. In what follows, we deal with linear velocity components normalized to the equatorial critical velocity $V_{\rm c}$ as

\begin{equation} 
\begin{array}{rcl}
\displaystyle v\sin i & = & \displaystyle u\sin i+\delta(i) \\ 
\displaystyle v & = & V_{\rm d}/V_{\rm c}; \ \ u = V_{\rm e}/V_{\rm c}; \ \ \delta(i) = \Delta V(\sin i)/V_{\rm c},
\label{eq26}
\end{array}
\end{equation}

\noindent where $\delta<0$ implies that the stellar surface rotation is accelerated from the pole toward the equator, $\delta>0$ indicates a rotation accelerated from the equator toward the pole, and $\delta=0$ is for rigid rotation.\par 
   To estimate the deviation $\delta=\delta(i)$ we used intrinsic Gaussian absorption lines, as in Sect.~\ref{rot_broad}, and calculated rotationally broadened flux line profiles with the GD relation put forward by \citet{elr11}. This GD relation is not adapted for surfaces undergoing differential rotation. However even a less detailed treatment of these effects is far out of the scope of the present statistical correction of rotational velocity distributions. These effects will be presented in detail in a series of forthcoming papers. \par
   Using the classical method of $V\!\sin i$ determination with the Fourier transform, we calculated $V_{\rm D}\!\sin i$ and $V_{\rm r}\!\sin i$ for a large space of parameters ($u,\alpha,\sin i$).\par
   The Fourier transformation of the gravity-darkened line profiles broadened by a surface differential rotation and rigid rotation was carried out with a rotational-broadening function produced by a linear limb-darkening law in a gray atmosphere of a rigid rotator, as is still currently used in the literature, where the first FT root $\sigma_1$ is related with the $V\!\sin i$ by the relation $\lambda(V\!\sin i/c)\sigma_1=0.66$. This was carried out to render similar in nature the model $V\!\sin i$ parameters to the stellar ones used in Paper I.\par
\begin{figure*}[]
\centerline{\includegraphics[scale=0.9]{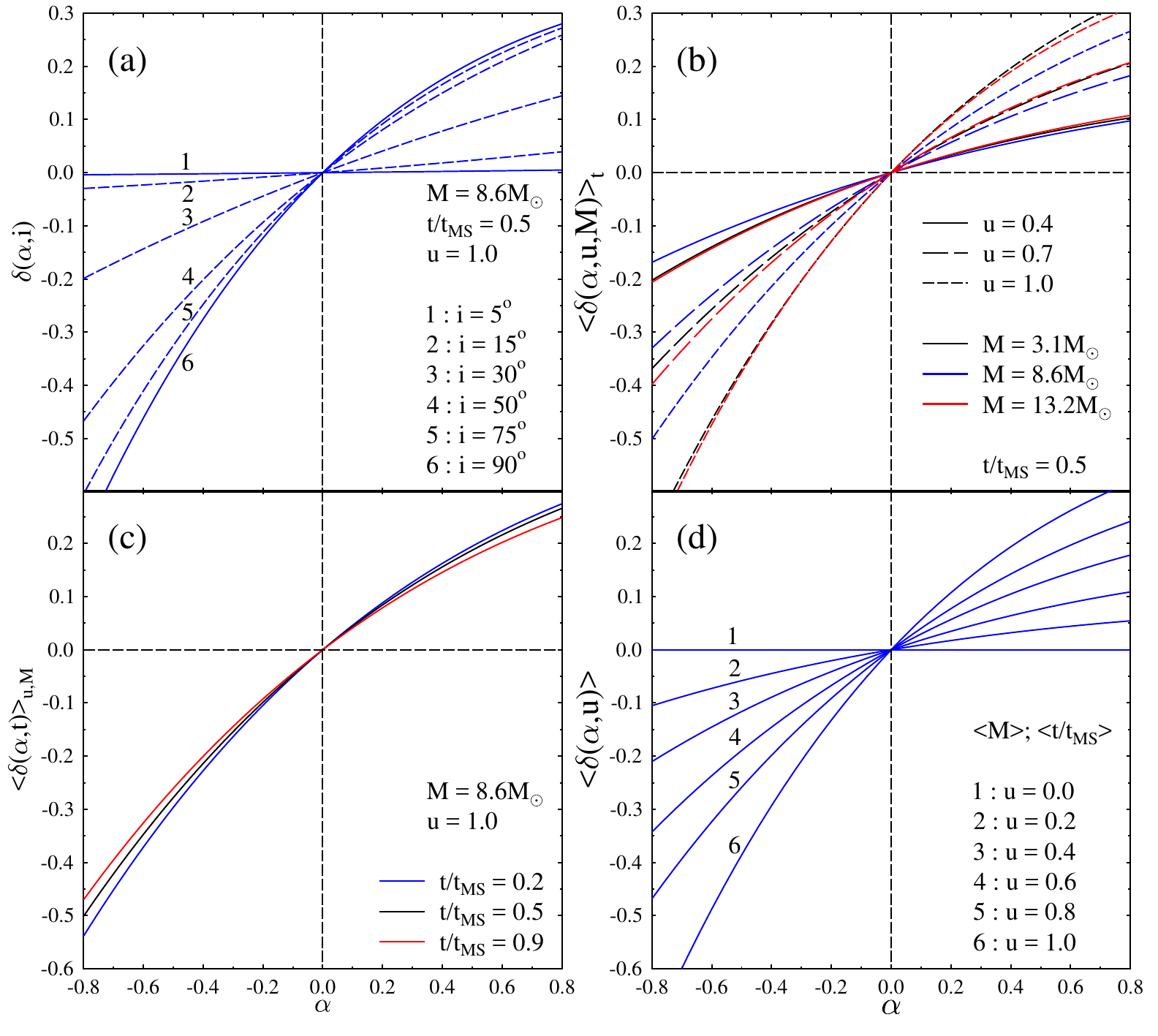}} 
\caption{\label{fig_16} (a) Deviation parameter $\delta(\alpha,u,i)$ as a function of $\alpha$ for $u=V_{\rm e}/V_{\rm c}=$ 1.0, mass $M/M_{\odot}=8.6,$ and fractional age $t/t_{MS}=$ 0.5, and for several inclination angles $5\degr<i<90\degr$. (b) Inclination-averaged deviation $\langle\alpha(\alpha,u)\rangle$ for velocity ratios $u=V_{\rm e}/V_{\rm c}=$ 0.4, 0.7, 1.0 in objects with masses $M/M_{\odot}=$ 3.1, 8.6, 13.2 at fractional age $t/t_{MS}=$ 0.5. (c) Inclination-averaged deviation $\langle\alpha(\alpha,t)\rangle$ for $M/M_{\odot}=$ 8.6 at fractional ages$t/t_{MS}=$ 0.2, 0.5, 0.9 rotating at $u=1.0$. (d) The adopted inclination-mass-age-averaged deviation $\langle\delta\rangle$ as a function of $\alpha$ for different values of the ratio $u$.} 
\end{figure*}

 
  Because the function $\Phi(u)$ is meant to represent the distribution of true-velocity ratios $u=V/V_{\rm c}$ obtained from inclination-angle averaged apparent velocity ratios, in what follows we have to use inclination-angle averaged deviations $\langle\delta\rangle$ that we define as follows: 

\begin{equation} 
\langle\delta(\alpha,u)\rangle = \frac{4}{\pi}\int_0^{\pi/2}\delta(\alpha,u,i)P(i)\,di,
\label{eq29}
\end{equation}

\noindent where $P(i)\,di\!=\!\sin i\,di$. In Fig.~\ref{fig_16}a are shown the deviations $\delta(\alpha,u,i)$ as a function of $\alpha$ for an object with mass $M/M_{\odot}\!=\!8.6$, fractional main-sequence age $t/t_{MS}\!=\!0.5$ ($t_{MS}$ is the time spent by a star in the main-sequence evolutionary phase) rotating with $u\!=\!V_{\rm e}/V_{\rm c}=\!1.0$, and for several inclination angles $5\degr\!<\!i\!<\!90\degr$. The $i-$averaged deviation $\langle\delta(\alpha,u)\rangle$ against $\alpha$ for masses $M/M_{\odot}=$ 3.1, 8.6, 13.2, which are all at the fractional age $t/t_{MS}\!=\!0.5$ and rotating at ratios $u\!=\!V_{\rm e}/V_{\rm c}\!=\!0.4$, 0.7, 1.0, is shown in Fig.~\ref{fig_16}b. Figure~\ref{fig_16}c shows the $i-$averaged $\langle\delta(\alpha,u)\rangle$ deviations obtained for $u=1$ in $M/M_{\odot}=$ 8.6 at three fractional ages $t/t_{MS}=0.5$, 0.5, and 0.9.\par
  Owing to the rather significant dependence of $\langle\delta(\alpha,u)\rangle$ with the stellar mass, we ought to study $\Phi(u)$ distributions obtained for several representative intervals of stellar masses. Unfortunately, the total number of objects in the entire sample is too low to define statistically reliable sub-groups. Having then a unique distribution $\Phi(u)$ for masses that are mostly in the interval $3.0\lesssim M/M_{\odot}\lesssim30.0$, we obtained $\langle\delta(\alpha,u,M)\rangle_t$ curves, similar to those shown in Fig.~\ref{fig_16}b, for the 11 mass-age stellar groups given in Table~\ref{tab_2}. The adopted deviation $\langle\delta(\alpha,u)\rangle$, shown in Fig.~\ref{fig_16}d and used in the final calculations, is the weighted average by the fraction of stars in each mass-age group.\par

 To simplify notations, in what follows we continue to use the notation $\delta$ for the aspect angle averaged deviation $\langle\delta(\alpha,u)\rangle$ and write  \par
 
\begin{equation}
\displaystyle  v = u+\delta(u,\alpha)  
\label{uvdelta}
.\end{equation}

 Since by definition $0\leq v\leq1$ and $0\leq u\leq1$, from Eq.~(\ref{uvdelta}), the theoretical range of possible values of $\delta$ is restricted to $-1\leq\delta\leq1$. Nevertheless, from the calculated line profiles it follows that $\lim_{\alpha\to+\infty}\delta(u,\alpha)=$ $\delta_{\rm L}(u)<1$ for all values of $u$. \par 

\subsubsection{Distributions of apparent and actual equatorial velocities ratios $V/V_{\rm c}$}\label{sect_23} 

   Following the principles for convolutions established in Appendix~D of Paper I, the relation between the observed distribution $\Psi(v)$ and the corrected distribution $\Phi(u)$, takes the form
 
\begin{equation}
\begin{array}{rcl}
\displaystyle \Psi(v) & = & \displaystyle \int_{u_1(v)}^{u_2(v)}\!\!\Phi(u)\pi(v-u)\,du \\ 
 & &  \\
\displaystyle \Pi(v-u) & = & \displaystyle [\phi(\alpha)/|\partial\delta/\partial\alpha|]_v.\\
\end{array}
\label{eq30}
\end{equation}  

\noindent The function $\Psi(v)$ is the distribution of ratios $v=V/V_{\rm c}$, where $V$ does not represent a actual equatorial velocity, but a kind of linear velocity averaged over the apparent stellar hemisphere that accounts for the observed line broadening. Conversely, the variable $u$ represents the ratio of actual true equatorial rotational velocities $V_{\rm eq}/V_{\rm c}$. In Eq.~(\ref{eq30}) the function $\phi(\alpha)$ is the probability for the occurrence of the differential rotation parameter $\alpha$. The lower limit of integration over $u$ in Eq.~(\ref{eq30}) is determined by the condition $u_1(v)=v-\delta_{\rm L}(u_1)$ as noted after the relation in Eq.~\ref{uvdelta}, while $u_2=1$ is the natural upper limit. Obviously, this convolution can also be understood as an integration over the interval of possible values of $\delta$ of the functional product $\Phi(v-\delta)\pi(\delta)$. \par    
  In Eq.~(\ref{eq30}) only $\Psi(v)$ is available, while neither of the functions $\Phi(u)$ and $\phi(\alpha)$ is known. Moreover, since $\delta(u,\alpha)$ depends on $u$ and $\alpha$, the functions $\Phi(u)$ and $\phi(\alpha)$) are mutually dependent. The problem is then to apply an iterative {\it blind deconvolution method} to obtain a solution for $\Phi(u)$ \citep[e.g.,][]{ayers_1988,tsum_1994,li_2016}. The application of such a method would require us to derive the probability function $\phi(\alpha),$ which obeys the constraining condition given by the relation $\delta=\delta(u,\alpha),$ and to deal with a distribution function $\pi(v-u)$ whose mode and dispersion change with $v$. Instead of discussing in detail the convergence of the method and the uniqueness of the possible inferred solutions for $\Phi(u)$ and $\phi(\alpha)$, which could lead us beyond the scope of the present section, we preferred to estimate only the order of magnitude of the deviations that might be expected to be induced by the differential rotation on the distributions of rotational velocity ratios $V/V_{\rm c}$, by using in Eq.~(\ref{eq30}) analytical expressions for $\Phi(u)$ and $\phi(\alpha)$ characterized by free parameters.\par  
  In Fig.~\ref{vsini_p} we see that the deviations produced on the $V\!\sin i$ values can probably be unrealistic if we impose differential rotation parameters $|\alpha|>1$. Hence, we assume simple functional forms for $\phi(\alpha)$ restricted to the interval $-1\leq\alpha\leq+1$ and characterized by a dispersion-like parameter $a_{\alpha}$ ($\sigma^2_{\alpha}\sim1/2a_{\alpha}$) and the mode-like parameter $\alpha_o$ as follows:
   
\begin{equation}
\begin{array}{rcl}
\phi(\alpha) & = & \phi_o(1-\alpha^2)\exp[-a_{\alpha}(\alpha-\alpha_o)^2] \\
\label{eqphia_assum}
\end{array}
,\end{equation}   

\noindent where $\phi_o$ is the normalization constant that depends on $a_{\alpha}$ and $\alpha_o$. We adopted $a_{\alpha}=4.5.0$ and 250.0 and several values for $\alpha_o$ from $-0.5$ to $+0.5$.\par
 The distribution of equatorial velocity ratios $\Phi(u_{\rm eq})$ [$u_{\rm eq}=V_{\rm eq}/V_{\rm c}$] is unknown by definition. The function $\Phi(u)$ obtained in Paper I for Be stars, after corrections for gravity darkening effect and microturbulent motions, must be considered to be, in principle, affected by the differential rotation of stars. It is then taken here as representing $\Psi(v)$. To remain, however, not far from the physical circumstances that determine the function $\Phi(u)$, for the present exercise we assume $\Phi(u_{\rm eq})$ given by an analytical expression with roughly the same mode and FWHM  as $\Phi(V/V_{\rm c})$ of Paper I mentioned above. So, we have

\begin{equation}
\begin{array}{rcl}
\Phi(u_{\rm eq}) & = & \Phi_ou^a_{\rm eq}\exp[-b(u_{\rm eq}-u_o)^2]; \ \ \ 0\leq u_{\rm eq} \leq1
\label{eqphiu_assum}
\end{array}
\end{equation}  

\noindent with $\Phi_o$ as the normalization constant, $a=0.649$, $b=29.94$ and $u_o=0.671$.\par
    
\begin{figure*}[] 
\centerline{\includegraphics[scale=1.0]{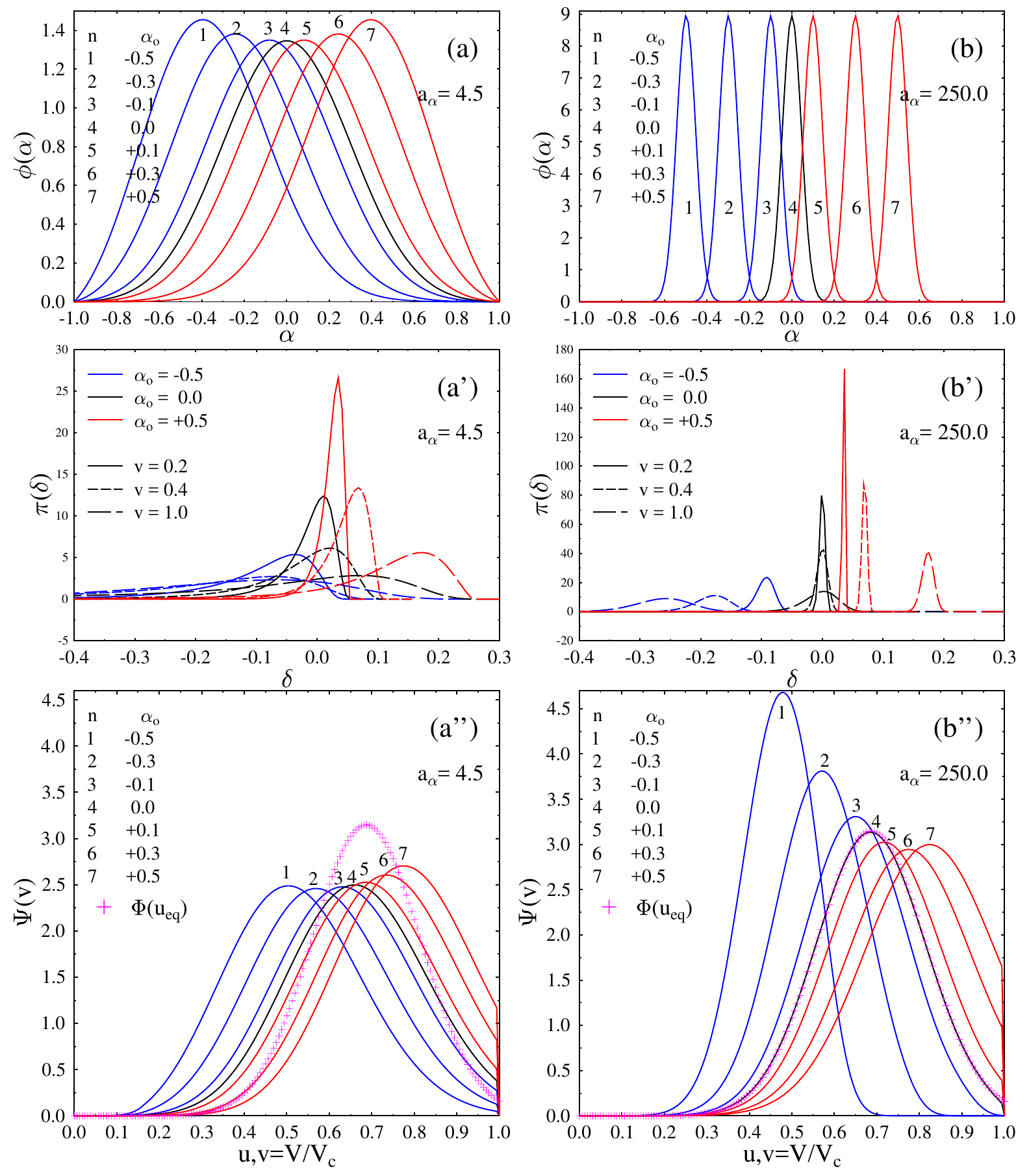}} 
\caption{\label{fig_18} (a) Assumed occurrence probability functions $\phi(\alpha)$ of the differential rotation parameter $\alpha$ with $a_{\alpha}=4.5$; $\alpha_o<0$ (blue curves), $\alpha_o=0$ (black curve), $\alpha_o>0$ (red curves). (b) Idem
as in (a), but for $a_{\alpha}=250.0$. (a') Occurrence probability functions $\pi(\delta)$ of the differential rotation parameter $\delta$ with $a_{\alpha}=4.5$ for different values of $\alpha_o$ and true velocity ratios $u$. (b) Idem
as in (a'), but for $a_{\alpha}=250.0$. (a'') Distributions $\Psi(v)$ of the equatorial velocity ratios $v=V/V_{\rm c}$, where $v$ is now affected by the differential rotation according to the distributions $\phi(\alpha)$ of the parameter $\alpha$ given in (a). (b'') Idem as in (a'), but for $\phi(\alpha)$ shown in (b). In (a'') and (b'') is also shown the assumed distribution  $\Phi(u_{\rm eq})$ (magenta crosses)}
\end{figure*}

  Figures~\ref{fig_18}a,b show the assumed functions $\phi(\alpha)$ used in this exercise according to the combination of parameters $a_{\alpha}$ and $\alpha_o$ given in the inserts. Figures.~\ref{fig_18}a',b'
reproduce some functions $\pi(\delta)=\pi(v-u)$ [see EQ.~(\ref{eq30})] for three values of $\alpha_o$ in $\phi(\alpha)$ and corresponding to three different values of $v$. Figures~\ref{fig_18}a",b" show the respective distributions $\Psi(v)$ calculated from Eq.~(\ref{eq30}). In these figures  the function $\Phi(u_{\rm eq})$ (red crosses), given by Eq.~(\ref{eqphiu_assum}), is also shown. To give a quantitative overview of the effects carried by the differential rotation on the distribution of ratios $u=V/V_{\rm c}$, Table~\ref{tab_mode} reproduces the modes of the functions $\phi(\alpha),$ the calculated distributions $\Psi(v)$, and the FWHM of the resulting `apparent' functions $\Psi(v)$.\par  
   
\begin{table}[]
\caption[]{\label{tab_2} Mass-age groups formed by the sample of Be stars entering the `observed'
distribution $\Phi(u)$.}
\tabcolsep 1.7pt
{\scriptsize \begin{tabular}{c|ccr|cccr|cccr} 
\hline
\noalign{\smallskip}
   & \multicolumn{3}{c|}{$t/t_{\rm MS}\leq0.4$} & & \multicolumn{3}{c|}{$0.4<t/t_{\rm MS}\leq0.8$} & &
\multicolumn{3}{c}{$t/t_{\rm MS}>0.8$} \\
\noalign{\smallskip}
\cline{2-4}\cline{6-8}\cline{10-12}
\noalign{\smallskip}
   & $\langle M/M_{\odot}\rangle$ & $\langle t/t_{\rm MS}\rangle$ & n && $\langle M/M_{\odot}\rangle$ & 
$\langle t/t_{\rm MS}\rangle$ & n &&  $\langle M/M_{\odot}\rangle$ & $\langle t/t_{\rm MS}\rangle$ & n \\
$M/M_{\odot}\leq7.5$    & \ \ 5.0 & 0.16 & 22 &&  \ \  5.0 & 0.65 & 25 &&  \ \  5.3 & 0.97 &  36 \\
$7.5<M/M_{\odot}\leq15$ & \ \ 9.4 & 0.19 & 17 &&      10.7 & 0.61 & 32 &&      10.5 & 0.95 &  52 \\
$15<M/M_{\odot}\leq25$  & 19.2 & 0.20 &  8 &&         19.8 & 0.72 &  7 &&      19.4 & 0.96 &  16 \\
\noalign{\smallskip}
\cline{2-12}
\noalign{\smallskip}
   & \multicolumn{3}{c|}{$t/t_{\rm MS}<0.5$} & & \multicolumn{3}{c|}{$t/t_{\rm MS}\geq0.5$} & &
\multicolumn{3}{c}{} \\   
\noalign{\smallskip}
\cline{2-4}\cline{6-8} 
\noalign{\smallskip}
   & $\langle M/M_{\odot}\rangle$ & $\langle t/t_{\rm MS}\rangle$ & n && $\langle M/M_{\odot}\rangle$ & 
$\langle t/t_{\rm MS}\rangle$ & n && \multicolumn{1}{c}{$\langle M/M_{\odot}\rangle$} & $\langle t/t_{\rm MS}\rangle$ & n \\
$M/M_{\odot}>25$        & 30.3 & 0.21 &  3 && 47.4 & 0.87 & 15 && \multicolumn{1}{c}{\it 12.4} & {\it 0.70} & {\it 233} \\
\hline
\noalign{\smallskip}
\multicolumn{12}{l}{Note: The global average mass and fractional age of the}\\
\multicolumn{12}{l}{sample are given in the bottom right box.}\\
\noalign{\smallskip}
\hline
\end{tabular}
}
\end{table}

\subsection{Comments on the results obtained}\label{comm_difrot} 

 From the results presented in Sect.~\ref{eff_vsini} we conclude that if the stars had their atmosphere in differential rotation, the parameters $V\!\sin i$ obtained with the currently used measurement methods would depend on the degree of differential rotation and sensitively differ from the values expected if they were rigid rotators. The difference is due to, on one hand, the geometry of the constant radial-velocity curves contributing to the individual monochromatic Doppler displacements that contribute to the broadening of a spectral line, and on the other hand, to the inadequacy of the measurement methods, which are not adapted to the physical circumstances imposed by the differential rotation. In stars rotating with an angular velocity accelerated from the pole toward the equator ($\alpha<0$), the measured parameter $V\!\sin i$ implies that $V<V_{\rm eq}$, where $V_{\rm eq}$ is the actual linear rotational velocity of the equator. Conversely, a differential rotation accelerated from the equator toward pole ($\alpha>0$) leads to a $V\!\sin i$ parameter where $V>V_{\rm eq}$. It also follows from these results that for the rotation law given in Eq.~\ref{eq27} and for a given absolute value $|\alpha|$, the differences $|V-V_{\rm eq}|$ are larger for $\alpha<0$ than for $\alpha>0$. The precise value of the difference $\Delta V=V-V_{\rm eq}$ depends on the specific differential rotation law, on the value of $V_{\rm eq}$, and on the spectral line used to carry out the measurements due to the different sensitivities to the physical conditions in their formation regions [see Sect.~\ref{rad_sf}, item 1)].\par 

  Regarding the distribution of velocity ratios $v=V/V_{\rm c}$, where $V$ is not the stellar equatorial velocity but accounts for the spectral line broadening as discussed in Sect.~\ref{sect_23}, Table~\ref{tab_mode} reveals that if the surface rotation laws are sketched with the relation in Eq.~\ref{eq27}, for $\alpha<0$ the modes of the distributions $\Psi(v)$ are $v_{\rm mode}<u^{\rm mode}_{\rm eq}$, and $v_{\rm mode}>u^{\rm mode}_{\rm eq}$ when $\alpha>0$. It also follows that for a given $|\alpha|$, the $|v_{\rm mode}-u^{\rm mode}_{\rm eq}|$ is larger when $\alpha>0$. On average, the FWHM of $\Psi(v)$ is an increasing function of $\alpha$. For a given value of $\alpha$, the larger the FWHM of $\Psi(v)$ is the smaller $a_{\alpha}$ is [larger `dispersion' of the probability function $\phi(\alpha)$]. When $a_{\alpha}\gtrsim4.5$ the FWHM is roughly uniform over a large interval of $\alpha_o$ values.\par
  To avoid confusion due to the notation used, we note $\Psi(v)_{\rm obs}$, the resulting distribution $\Phi(u)$  in Paper I, after corrections for gravity darkening and macroturbulent motions. The distributions $\Psi(v)$ issued from Eq.~(\ref{eq30}) are thus meant to be different cases of $\Psi(v)_{\rm obs}$. So, because of the displacement of the distribution of apparent velocity ratios caused by the differential rotation with respect to the distribution of the actual equatorial velocity ratios, we can expect that the number of Be stars with equatorial velocities attaining $V_{\rm eq}\gtrsim0.9V_{\rm c}$ can be larger than suggested by $\Psi(v)_{\rm obs}$ if on average the stars have rotation laws with $\alpha<0$. On the contrary, the number of rotators with $V_{\rm eq}\gtrsim0.9V_{\rm c}$ could be lower if they have rotation laws with $\alpha>0$. Finally, if our program Be stars rotate with $\alpha\simeq-0.3$ as the Sun, or $\alpha\simeq-0.2$ as predicted for stars with masses $2\lesssim M/M_{\odot}\lesssim4$ by \citet{rieut13}, the $\Psi(v)_{\rm obs}$ will have its mode displaced toward lower values by $\Delta v\simeq-0.07\pm0.01$ or $\Delta v\simeq-0.11\pm0.01$, respectively, as compared with the mode of the distribution of their actual equatorial ratios $u_{\rm eq}=V_{\rm eq}/V_{\rm c}$; the noted uncertainty accounts for the two values of $a_{\alpha}$ used here.\par   
 
\begin{table}[h!]
\centering
\caption[]{\label{tab_mode} Modes of functions $\phi(\alpha)$ and modes and FWHM of distributions $\Psi(v)$}
\tabcolsep 5.0pt
\begin{tabular}{c|ccccccc}
\hline
\noalign{\smallskip} 
 & \multicolumn{3}{c}{$a_{\alpha}=4.5$} & & \multicolumn{3}{c}{$a_{\alpha}=250.0$} \\
\noalign{\smallskip}
\cline{2-4}\cline{6-8} 
\noalign{\smallskip}
 $\alpha_o$ & $\alpha_{\rm mode}$ & $v_{\rm mode}$ & FWHM && 
$\alpha_{\rm mode}$ & $v_{\rm mode}$ & FWHM \\
  -0.5 & -0.400 &  0.506 &  0.384 && -0.500 & 0.481 &  0.206 \\ 
  -0.4 & -0.325 &  0.538 &  0.387 && -0.400 & 0.525 &  0.230 \\
  -0.3 & -0.237 &  0.569 &  0.388 && -0.300 & 0.569 &  0.251 \\
  -0.2 & -0.162 &  0.600 &  0.388 && -0.200 & 0.613 &  0.270 \\
  -0.1 & -0.087 &  0.631 &  0.388 && -0.100 & 0.650 &  0.287 \\
   0.0 &  0.000 &  0.656 &  0.388 &&  0.000 & 0.688 &  0.302 \\
   0.1 &  0.088 &  0.688 &  0.388 &&  0.100 & 0.719 &  0.317 \\
   0.2 &  0.163 &  0.713 &  0.387 &&  0.200 & 0.750 &  0.330 \\
   0.3 &  0.238 &  0.738 &  0.387 &&  0.300 & 0.775 &  0.342 \\
   0.4 &  0.325 &  0.756 &  0.387 &&  0.400 & 0.800 &  0.353 \\
   0.5 &  0.400 &  0.775 &  0.387 &&  0.500 & 0.825 &  0.363 \\
\noalign{\smallskip} 
\hline
                   &&&&&& $u_{\rm eq}^{\rm mode}$ & FWHM \\
$\Phi(u_{\rm eq})$ &&&&&& 0.688 &  0.302 \\
\noalign{\smallskip} 
\hline
\end{tabular}
\end{table} 
   
\section{Some hidden effects on the $V\!\sin i$ parameter}\label{uncert_vsini} 

  A critical study of the distribution of the $V\!\sin i$ parameter and the conclusions that could be drawn from it cannot be completed if, nevertheless,  we do not mention a number of conceptual and measurement uncertainties related to the determination of this quantity, other than differential rotation, which can also render the study of the true rotational properties of stars difficult.\par 
  From the existing compilations of Be-star $V\!\sin i$ parameters, we realize that the typical dispersions of independent estimations for the same objects carried out by different authors are typically $20\lesssim1\sigma_{V\!\sin i}\lesssim50$ km~s$^{-1}$ \citep{slett82,chauv01,yud01,gleb05}. On the other hand, more or less regular and homogeneous followup observations of Be stars reveal variations of $V\!\sin i$ on the order of 30 km~s$^{-1}$ \citep{chauv01,rivi13}. Deviations can sometimes be as high as 100 to 150 km~s$^{-1}$ as for HD 45910, HD 52721, and HD 135734 \citep{yud01}, or they can even attain 200 km~s$^{-1}$ as in $\gamma$~Cas \citep{slett82,chauv01,harmc02}. Such discrepancies are possibly not only related to the method used to determine the $V\!\sin i$, but probably also with physical phenomena that affect the spectral line formation.\par
  Apart from macroturbulent motions evoked in Sect.~4 of Paper I, which pertain to  low $V\!\sin i$ ($V\!\sin i\lesssim150$ km~s$^{-1}$), the $V\!\sin i$ of Be stars and active stars, in general, are sensitive to spectral line variation and broadening owing to a sum of physical phenomena that can contribute individually with systematic and/or irregular components of a few tenths of km~s$^{-1}$. In some cases these stars can significantly  blur or obliterate the information on the stellar rotation carried by the $V\!\sin i$. Among such phenomena and/or uncertainties, the following deserve some attention: \par

\begin{itemize}
\item[$i$)]   the angular momentum content in stars;  
\item[$ii$)]  double-valued $V\!\sin i;$ 
\item[$iii$)] gravity darkening effect on the radiation source function;
\item[$iv$)]  effects carried by intrinsically asymmetric rotational broadening functions; 
\item[$v$)]   deviations produced by expanding layers; 
\item[$vi$)]   deformations of line profiles due to tidal interactions in binary systems;
\item[$vii$)] presence of circumstellar envelopes or discs (CD).  
\end{itemize}  
 To this list we should add: \par 
\begin{itemize}
\item[$viii$)] wavelength-dependent limb-darkening within a spectral line \citep{undh68,stomi73,coll95,levh14};
\item[$ix$)]   uncertainties due to nonradial pulsations \citep{aerts1,aerts2}. 
\end{itemize}  
 
\noindent These uncertainties were discussed by the cited authors and the specific publications  provide details. \par

\subsection{Angular momentum content in stars}\label{cont_angmom}

  The most frequent descriptions of rotating objects are based on stellar models that start evolving in the ZAMS as rigid rotators. Evolution then  rapidly brings about the angular momentum redistribution so that stars become differential rotators \citep{maed09}. Resuming the angular momentum content in a star with the ratio $\tau_{\Omega} = K/|W|$, where $K$ is its total rotational kinetic energy and $W$ represents its gravitational potential energy, these models are characterized by ratios

\begin{equation}
0 \leq \tau_{\Omega} \lesssim 0.01, 
\label{eq33}
\end{equation}

\noindent where $\tau_{\Omega}\approx0.01$ identifies the rigid critical rotators. At this limit, owing to the mass-compensation effect, the bolometric luminosity produced in the stellar core can decrease by no more than some 5\% with respect to its value in stars at rest having masses $M/M_{\odot}\simeq60$, and to about 10\% for $M/M_{\odot}\simeq1.5$ \citep{boden71,mae00,ekst08}. However, secularly stable axisymmetric models can be constructed with energy ratios as large as 

\begin{equation}
0.01 \lesssim \tau_{\Omega} \lesssim 0.10 
\label{eq34}
,\end{equation} 

\noindent which necessarily correspond to diffe\-ren\-tial rotators in depth and where the reduction of the core bolometric luminosity at $\tau_{\Omega}\sim0.10$ attains from 40\% in stars with $M/M_{\odot}\simeq60$ to 80\% in those with $M/M_{\odot}\simeq1.5$ \citep{boden71,clem79,erimu85,deup01,jack05}.\par 
 Except for the Sun and perhaps KIC~10526294 \citep{tria15}, where an upper limit could be attempted for $\tau_{\Omega}$, at the moment it is impossible to produce systematic estimates of this ratio in other stars. However, if actual Be stars and other massive rapid rotators had ratios $\tau_{\Omega}$ as high as in Eq.~(\ref{eq34}), significant difficulties could appear to match the right distributions of surface effective temperatures and gravities to predict correctly  profiles of spectral lines that enable us to obtain consistent values of $V\!\sin i$. Needless to say, the diagnostic of stellar masses and ages would also be uncertain if models of rotating stars did not correspond to the right rotation-energy content in the analyzed stars \citep{zor86,cosmi85,love06,gill08,deup14a,deup14b}.\par
 
\begin{figure}[]
\centerline{\includegraphics[width=6.5truecm,height=6.5truecm]{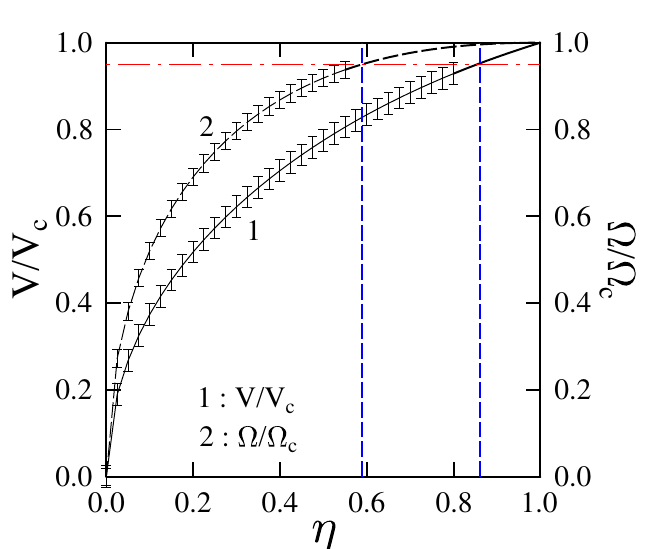}} 
\caption{\label{omet} Angular velocity ratio $\Omega/\Omega_{\rm c}$ as a function of 
$\eta$, where the uncertainty bars indicate differences for masses and ages in the main-sequence phase. It is shown that $\Omega/\Omega_{\rm c}\simeq0.95$ (red line) identifies the force ratio $\eta\simeq0.6$ at the equator, which implies a 40\% under-critical rotation. The ratio $V/V_{\rm c}\simeq0.95$ (red line) corresponds to $\eta\simeq0.86$.} 
\end{figure}

\subsection{Double-valued $V\!\sin i$}\label{dvvsi}
  
   The geometry of the stellar surface depends strongly on its surface rotation law. Objects with laws accelerated toward the pole may present polar dimples and, thus, the largest contribution to the Doppler broadening of spectral lines comes from a point of  intermediate latitude situated on the stellar meridian facing the observer. For large inclination angles, the contribution to the line broadening produced by the closed curves corresponding to $\mathscr{C}>1$ (see Fig.~\ref{fig_14}c) can be partially hidden in the troughs, which in addition can be more or less dimmed by the limb darkening. The relation between the FWHM of spectral lines and the $V\!\sin i$ then becomes double valued, as shown in \citet{zor86,zor94,zord04}.\par

\begin{figure}[]
\center\includegraphics[scale=1.0]{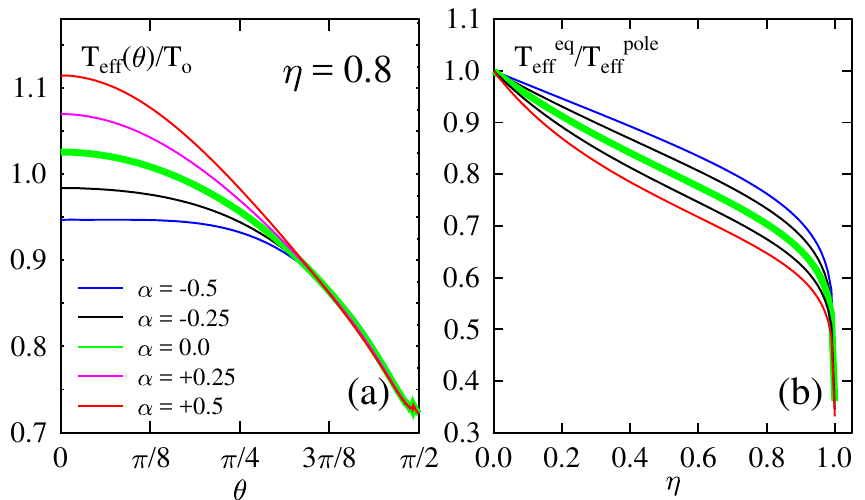} 
\caption{\label{tefep} (a) Effective temperature ratio $T_{\rm eff}/T_o$ against the colatitude $\theta$ derived by \citet{zor16b} using Eq.~(\ref{eq35b}) for $\eta=0.8$ and several values of the differential rotation parameter $\alpha$. The paramater $T_o$ is the effective temperature of the stellar counterpart without rotation. (b) Ratio $T^{\rm eq}_{\rm eff}/T^{\rm pole}$ against $\eta$ and for the same values of $\alpha$ as in (a).}
\centerline{\includegraphics[]{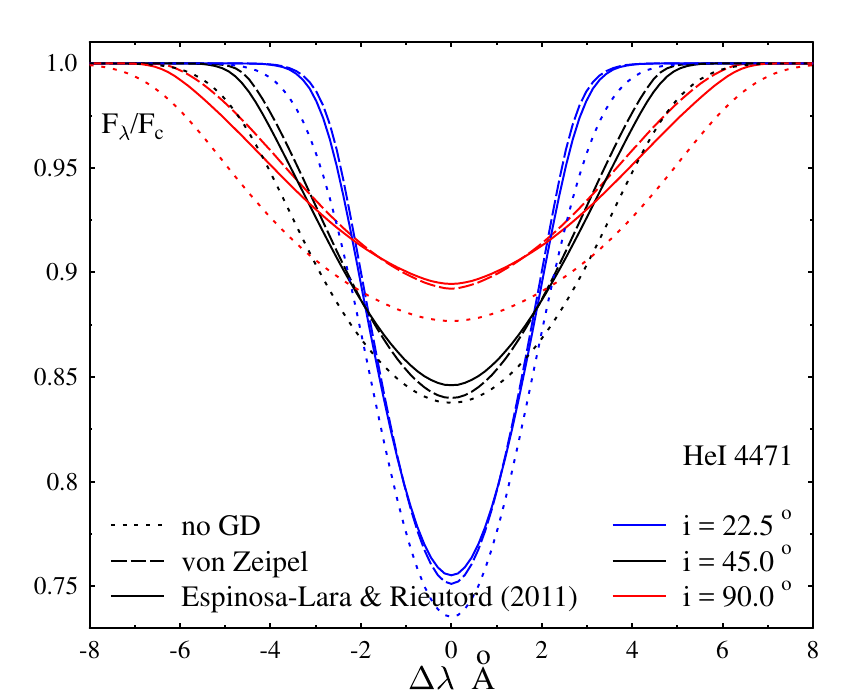}} 
\caption{\label{lin_elr} Rotationally broadened Gaussian line (assimilated to \ion{He}{i}\,4471) by a model star in rigid rotation with $\eta=0.9$ and $pnrc$ parameters $T_{\rm eff}=$ 16\,000 K and $\log g=$ 4.0. The dotted lines indicate no GD effect; the dashed lines indicate a GD effect using the von Zeipel formulation with $\beta_1=1.0$; and the solid line indicates the GD effect using Espinosa-Lara \& Rieutord's (2011) formulation. The colors indicate the inclination angle $i$.} 
\end{figure}

\subsection{Comments on the gravity darkening}\label{comm_gd} 

\subsubsection{Uncertainties related to the gravity darkening exponent}\label{gd_exp}

  The upper layers of the envelope of massive and intermediate-mass stars are globally in hydrostatic and radiative equilibrium. Accordingly, a relation between the emerging bolometric radiation flux $F_{\rm bol}$ at a given colatitude $\theta$ and the corresponding effective gravity $g_{\rm eff}(\theta)$ takes the form \citep{elr11,rieu15,zor16b}

\begin{equation}
F_{\rm bol}(\theta) = \sigma_{\rm SB}T^4_{\rm eff}(\theta) = f(\eta,\theta)g_{\rm eff}(\theta)
\label{eq35}
,\end{equation}

\noindent where $f(\eta,\theta)$ is a function of the rotation law in the stellar atmosphere and of the equatorial rotational rate $\eta$ defined in Eq.~(\ref{eq29-1}). However, the GD relation is still frequently written as
 
\begin{equation}
\sigma_{\rm SB}T^4_{\rm eff}(\theta) = \kappa(\eta)g^{\beta_1}_{\rm eff}(\theta),
\label{eq35b}
\end{equation}    
 
\noindent where the coefficient $\kappa(\eta)$ and the exponent $\beta_1$ are considered independent of the stellar colatitude $\theta$. Actually, if radiative equilibrium is imposed in the atmosphere and conservative rotation laws, in particular the rigid rotation, it can be shown that $\beta_1=1$ \citep{vzpl24,tass78}. Otherwise, conserving the expression given in Eq.~(\ref{eq35b}) and the above assumption of constancy for $\kappa(\eta)$ and $\beta_1$, it follows that for non-conservative laws the GD exponent becomes $\beta_1\leq1$ \citep{kippn77,mae99,clar12}. In general, the two-dimensional nature of the radiative transfer in geometrically deformed stars (or three-dimensional as for rapidly rotating components in binary systems) induces a horizontal diffusion of light \citep{osak66}. Accordingly, the GD effect decreases the higher the stellar oblateness, so that $\beta_1\leq1$ or even $\beta_1\to0$ \citep{pust70,smith74,hadr92}.\par  

\begin{figure}[] 
\center\includegraphics[scale=1.0]{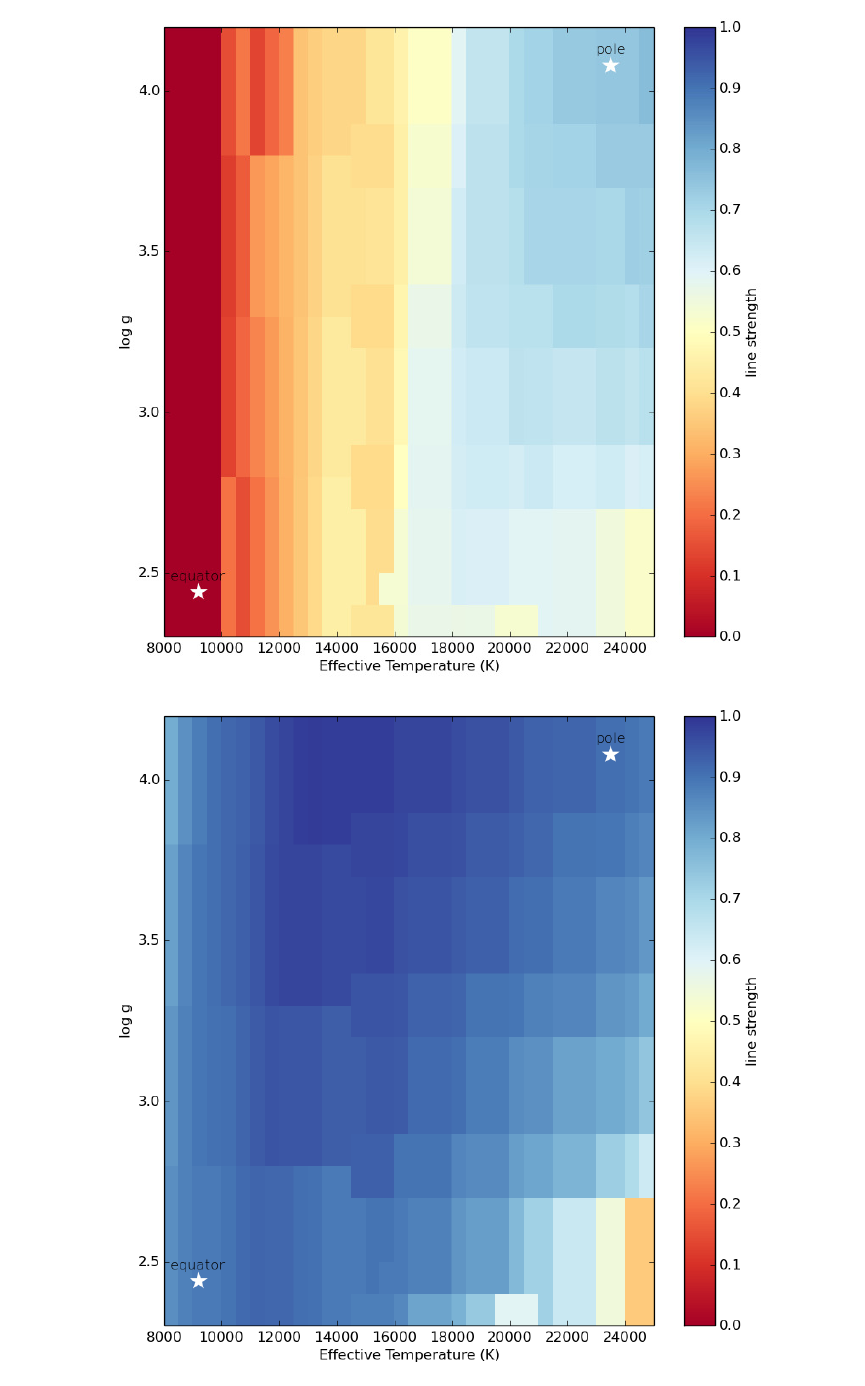}
\caption{\label{hemg_sens} Upper panel: Diagram showing the degree (in terms of relative equivalent width) of local contribution to the formation of the \ion{He}{i}\,4471 line in a rigid critical rotator. Lower panel: Degree of local contribution to the formation of the \ion{Mg}{i}\,4481 line. The white stars identify the pole and the equator.} 
\end{figure}
 
   The interferometric imaging of stellar surfaces with models of rapid rotators based on Eq.~(\ref{eq35b}), where $\kappa(\eta)$ and $\beta_1$ are kept constant, produces $\beta_1\lesssim1.0$ with $\beta_1$ decreasing as the flattening increases \citep{monn07,zhao09,che11,monn12,omar13,souz14}. However, \citet{elr11} showed that even in rigid rotators $\beta_1$ decreases as the $\eta$ increases and it is also a function of the colatitude $\theta$.  The dependence with $\theta$ implies that the GD exponent determined empirically is necessarily inclination angle $i-$dependent \citep{rieu15,zor16b}. The information that $\beta_1$ might carry on the rotation of stars is then somewhat obliterated, which is similar to what happens with the $V\!\sin i$ parameter due to the unknown $\sin i$.\par 
   According to relation in Eq.~(\ref{eq35}) the variation of the effective temperature with $\theta$ is somewhat different as obtained from Eq.~(\ref{eq35b}). This difference is even stronger if the stellar surfaces undergo differential rotation as shown in \citet{zor16b}. Figure~\ref{tefep}a shows the variation of the ratio of effective temperatures  $T_{\rm eff}/T_o$ obtained in \citet{zor16b} ($T_o$ is the effective temperature of the non-rotating stellar counterpart) against the colatitude $\theta$ in differentially rotating models with equatorial force ratio $\eta=0.8$ and several values of the parameter $\alpha$ in $\Omega(\theta,\alpha)$ given by Eq.~(\ref{eq27}). Figure~\ref{tefep}b shows the ratio $T^{\rm eq}_{\rm eff}/T^{\rm pole}$ against $\eta$ and for the same values of $\alpha$ as in Fig.~\ref{tefep}a. In these figures, the closest variations of the effective temperature with $\theta$ and $\eta$ to the case described by Eq.~(\ref{eq35b}) with $\beta_1=1.0$ are those depicted by the green curves, which correspond to $\alpha=0$.\par    
   Differences in the predicted line profiles, in particular their equivalent widths, can appear according to the theory used to describe the GD effect. Figure~\ref{lin_elr} shows rotationally broadened flux profiles of a Gaussian line whose central intensity and equivalent width respond to the effective temperature and gravity as does the \ion{He}{i}\,4471 line. These profiles were calculated using in turn Eq.~(\ref{eq35b}) with $\kappa=$ constant and $\beta_1=1$ of the classical formulation of the GD effect, and the  expression in Espinosa-Lara \& Rieutord (2011) given by  Eq.~(\ref{eq35}) for $\alpha=0$. As noted above, when $\alpha=0$, the $\theta-$dependent effective temperature does not differ greatly on which of these two GD theories is used. According to Fig.~\ref{tefep} much larger differences are expected when $\alpha\neq0$.\par  

\begin{figure*}[] 
\center\includegraphics[scale=0.85]{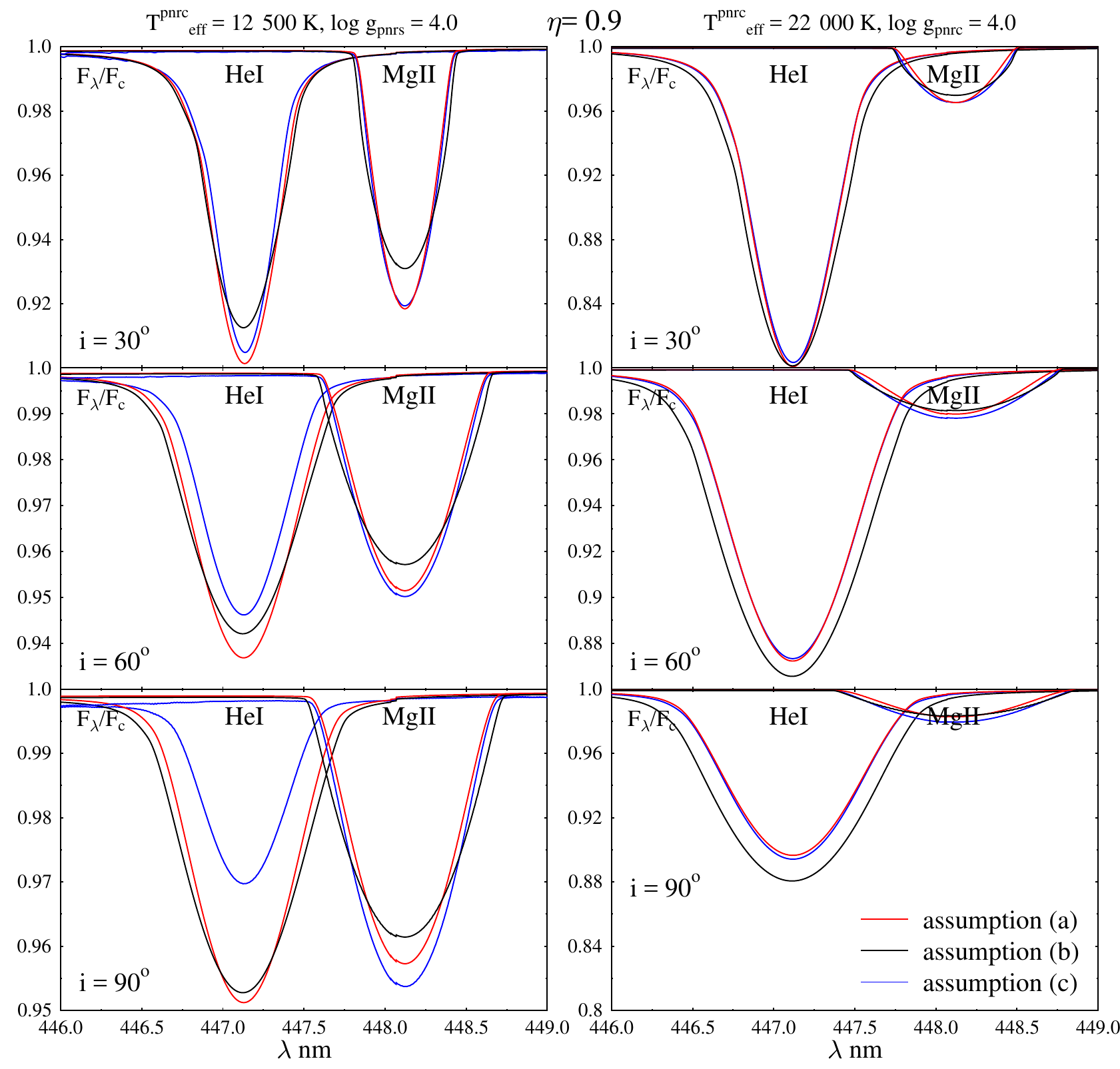}  
\caption{\label{fig_19} Rotationally broadened \ion{He}{i}\,4471 and \ion{Mg}{ii} 44781 line profiles produced by geometrically deformed model stars characterized by $(T^{\rm pnrc}_{\rm eff}=12\,500 K,\log g_{\rm pnrc}=4.0)$ and $(T^{\rm pnrc}_{\rm eff}=22\,000 K,\log g_{\rm pnrc}=4.0)$ and rotating at $\eta=0.9$. Red lines indicate hybrid gravity darkening with local emerging bolometric fluxes modulated according to Eq.~(\ref{eq35}), $\theta-$dependent effective gravity and local source functions determined by a uniform electron temperature $T_{\rm e}=\langle T_{\rm e}\rangle$ (assumption ``a"). Black lines indicate no gravity darkening and $\theta-$dependent effective gravity (assumption ``b"). Blue lines indicate classical gravity darkening with $\beta=1$ and the local source functions are determined by $T_{\rm e}[T_{\rm eff}(\theta),\log g(\theta)]$ (assumption ``c").}
\end{figure*}

\begin{table*}[]
\centering
\caption[]{\label{tab_app} Apparent fundamental parameters deduced according to se\-veral 
assumptions about the structure of gravity darkened atmospheres.}
\begin{tabular}{ccccccccccccccc}
\hline
\noalign{\smallskip} 
& & \multicolumn{6}{c}{$T^{\rm pnrc}_{\rm eff}=12\,500$~K} && \multicolumn{6}{c}{$T^{\rm pnrc}_{\rm 
eff}=22\,000$~K} \\
 & & \multicolumn{6}{c}{$\log g^{\rm pnrc}=4.0$~dex} && \multicolumn{6}{c}{$\log g^{\rm pnrc}=4.0$~dex} \\
 & & \multicolumn{6}{c}{$V_{\rm eq}=376$ km\,s$^{-1}$} && \multicolumn{6}{c}{$V_{\rm eq}=476$ km\,s$^{-1}$} \\
\noalign{\smallskip}
\cline{3-8}\cline{10-15} 
\noalign{\smallskip}
 $i$ & model & $W_{4471}$ & $W_{4481}$ & $T^{\rm app}_{\rm eff}$ & $\log g^{\rm app}$ & \multicolumn{2}{c}{$(V\!\sin i)^{\rm app}$} && $W_{4471}$ & $W_{4481}$ &
 $T^{\rm app}_{\rm eff}$ & $\log g^{\rm app}$ & \multicolumn{2}{c}{$(V\!\sin i)^{\rm app}$} \\
\cline{7-8}\cline{14-15} 
 & & $\AA$ & $\AA$ & K & dex & He & Mg && $\AA$ & $\AA$ & K & dex & He & Mg \\
\noalign{\smallskip}
       & (a) & 0.535 & 0.322 & 12\,850 & 4.10 & 174 & 171 && 1.273 & 0.161 & 22\,000 & 3.81 & 225 & 212 \\
30\degr & (b) & 0.523 & 0.330 & 12\,770 & 4.24 & 199 & 212 && 1.440 & 0.167 & 22\,000 & 4.07 & 260 & 264 \\
       & (c) & 0.501 & 0.349 & 12\,620 & 4.56 & 152 & 181 && 1.265 & 0.176 & 20\,830 & 3.71 & 223 & 239 \\
\noalign{\smallskip}
       & (a) & 0.472 & 0.311 & 12\,410 & 3.65 & 274 & 278 && 1.138 & 0.153 & 22\,210 & 3.62 & 348 & 348 \\
60\degr & (b) & 0.483 & 0.325 & 12\,490 & 3.98 & 311 & 334 && 1.382 & 0.163 & 22\,050 & 3.99 & 411 & 414 \\
       & (c) & 0.388 & 0.355 & 11\,730 & 4.22 & 231 & 293 && 1.148 & 0.178 & 20\,370 & 3.48 & 351 & 384 \\
\noalign{\smallskip}
       & (a) & 0.380 & 0.296 & 11\,640 & 2.73 & 289 & 301 && 0.974 & 0.143 & 22\,250 & 3.35 & 374 & 379 \\
90\degr & (b) & 0.416 & 0.318 & 12\,020 & 3.60 & 332 & 363 && 1.313 & 0.159 & 22\,340 & 3.90 & 446 & 450 \\
       & (c) & 0.251 & 0.364 &  9\,720 & 3.32 & 234 & 314 && 1.009 & 0.180 & 18\,580 & 3.20 & 375 & 415 \\
\noalign{\smallskip} 
\hline
\multicolumn{15}{l}{Note: (a) : GD models with uniform radiation source function; (b) : no GD but $\theta-$dependent effective gravity; (c): the classical}\\
\multicolumn{15}{l}{von Zeipel GD. $W_{4471}$ and $W_{4481}$ are the equivalent widths of the \ion{He}{i}\,4471 and \ion{Mg}{ii}\,4481 lines, respectively.}\\
\noalign{\smallskip}
\hline
\end{tabular}
\end{table*} 
    
\begin{figure*}[]
\center\includegraphics[scale=0.75]{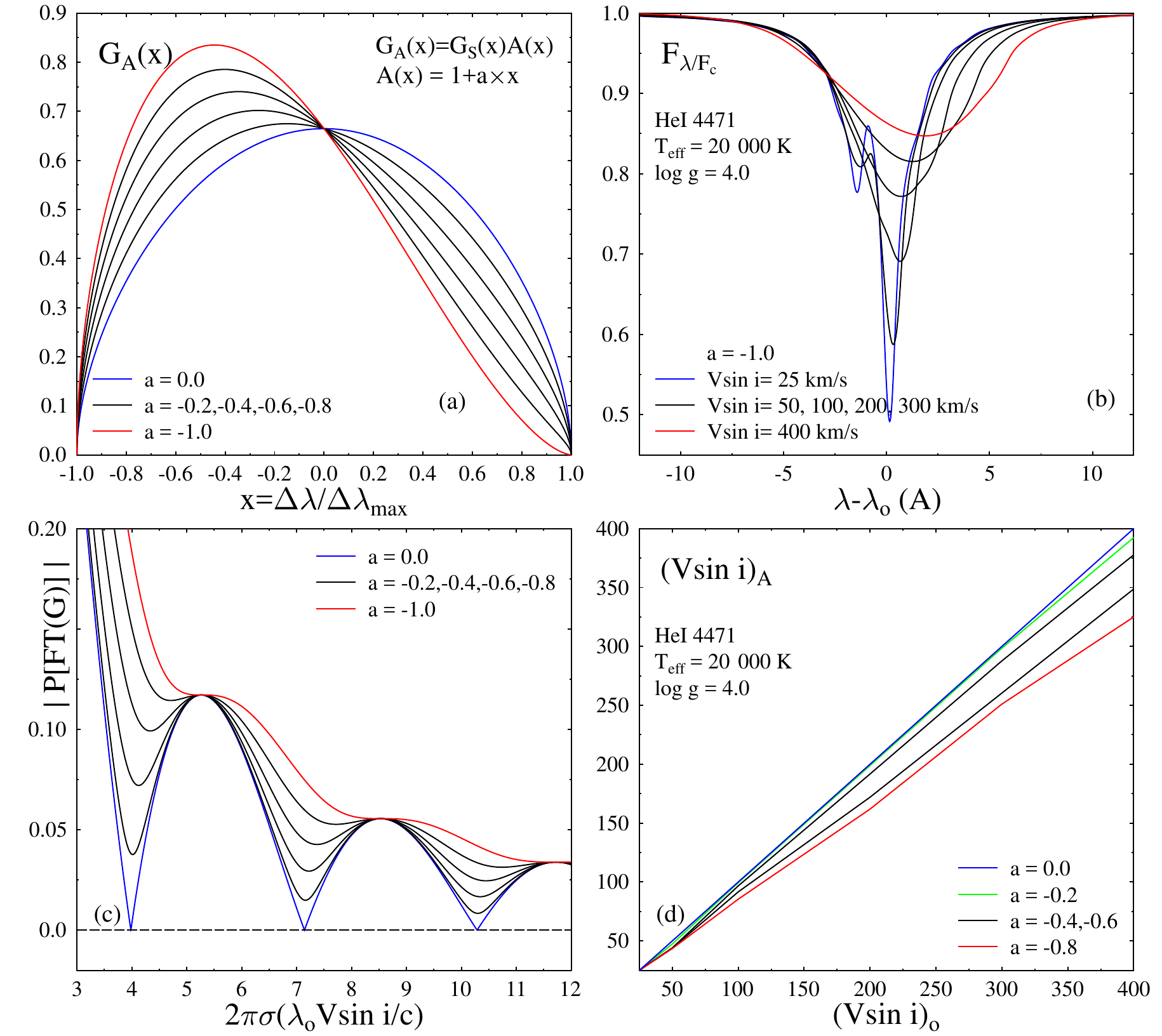}  
\caption{\label{fig_20} (a) Tested asymmetric rotation broadening functions (ARBF). (b) Broadened \ion{He}{i}\,4471 line profiles with ARBF having $a=-1$ and for several input $V\!\sin i$ parameters. (c) FT of the ARBF. (d) Resulting $(V\!\sin i)_{\rm A}$ parameters produced by ARBF against the model input $(V\!\sin i)_o$ [the colors correspond to the  distortion parameters $a$ in a)].}  
\end{figure*}

\subsubsection{Radiation source function in a gravity-darkened atmosphere}\label{rad_sf}

  To present the effects we want to stress in this section clearly, two types of phenomena deserve a comment: 
1) the sensitivity of line source functions to local physical conditions; and 2) the thermal structure on depth of stellar atmospheres as a function of the colatitude $\theta$ and its incidence on the line source function.\par 

\medskip
 1) {\sl Sensitivity of line source functions}\par 
\medskip
  From the classical theory of stellar atmospheres, it is well known that due to non-LTE effects the source function 
of spectral lines have selective sensitivities to collisional and radiative processes, which dominate the population of atomic levels; these depend on the particular structure of ions \citep{thom65,thom83,mih78}. Since a gravity-darkened
atmosphere in a rapidly rotating star displays a wide range of electron temperatures and densities over the hemisphere
projected toward the observer, spectral lines with different ``reactivities" to local formation conditions do not reflect in the same way, or with the same strength, the physical properties in a given region of this hemisphere.  To illustrate this effect, we calculated the $T_{\rm eff},\log g_{\rm eff}-$diagrams show in Fig.~\ref{hemg_sens},
where the color scaling corresponds to the relative equivalent width of the \ion{He}{i}\,4471 and \ion{Mg}{ii}\,4481 lines in a rigid critical rotator. We assumed $T^{\rm pole}_{\rm eff}=$ 24\,000 K and the color scale
ranges from 0, for no contribution, to 1, or maximum contribution to the central flux. The polar regions are situated in the upper right corner, while the equatorial regions are in the lower left corner. From these diagrams we can conclude that a gravity-darkened atmosphere contributes to the rotational broadening of the \ion{He}{i}\,4471 line likely in the upper latitudes, while the rotational broadening of the \ion{Mg}{ii}\,4481 line is produced by the entire hemisphere. Owing to these differences, spectra of rotating stars interpreted with models for non-rotating atmospheres can produce not only different values of apparent $(T_{\rm eff},\log g_{\rm eff})$-parameters, but also they lead to
different estimates of the $V\!\sin i$, as shown in the next subsection. The contribution of local specific intensities to the observed flux in a spectral line is still modulated by the local continuum flux, which in a gravity darkened atmosphere can be lower in the equatorial region than in other latitudes.\par 

\medskip
 2) {\sl The source function at $\tau_{\lambda}=2/3$}\par 
\medskip
    
  It was recognized very early that rotation and simultaneous hydrostatic and radiative equilibrium contradict each other, producing non-vanishing $\theta$-dependent divergence of radiation flux \citep{osak66}. Moreover, because stars are probably baroclinic (non-conservative rotation laws) their isobars and isopycnic surfaces are neither parallel nor isothermal, which makes even more difficult to know the actual behavior of $S_{\lambda}$ with $\theta$ at $\tau_{\lambda}=2/3$ , which is considered as the formation region of the radiation observed at $\lambda$. Unfortunately, up to now there are no detailed predictions made of atmospheric thermal structures as a function of $\theta$ in rapidly rotating stars. The current approximation used to model the spectra emitted by rotating objects is to assume that at each colatitude $\theta$ the stars have an internal atmospheric structure similar to that of a classical plan-parallel model-atmosphere characterized by the local parameters $[T_{\rm eff}(\theta),\log g_{\rm eff}(\theta)]$, where $T_{\rm eff}(\theta)$ is inferred with some GD law among those cited in Sect.~\ref{gd_exp}. Such a thermal structure has never been demonstrated to be consistent. So, at the moment, we can only simulate the effects produced by several extreme physical circumstances to see what the expected changes in the spectra are:\par  
\begin{itemize}
\item[(a)] {\sl Uniform source function}: The source function $S_{\lambda}$ in the radiation formation region is assumed to be the same for all $\theta$ and determined by the classic atmospheric temperature structure, whose effective temperature $T_{\rm eff}$ is the $\theta-$averaged effective temperature $\langle T_{\rm eff}\rangle$, where each local $T_{\rm eff}(\theta)$ is determined with Eq.~(\ref{eq35}). Each elementary atmosphere responds to the $\theta-$dependent surface effective gravity $g=$ $g_{\rm eff}(\theta)$. However, the amount of the emerging $\lambda-$dependent specific intensity at each $\theta$ is assumed to be modulated according to the bolometric flux given in Eq.~(\ref{eq35}).
\item[(b)] {\sl No gravitational darkening}: Following the predictions by \citet{pust70} and \citet{hadr92} that in rapid rotators $\beta_1\to0$, we assume that the studied spectral characteristics are produced by a non-spherical rotating star with no GD, where the internal structure of each local elementary atmosphere conforms with a uniform $T_{\rm eff}=$ $T^o_{\rm eff}$, but has a $\theta-$dependent $\log g$. The main difference with case (a) is that the local emerging specific intensity is not modulated by the bolometric flux given in Eq.~(\ref{eq35}) and that $T^o_{\rm eff}\neq\langle T_{\rm eff}\rangle$. 
\item[(c)] {\it Classical gravitational darkening}: We adopt the commonly used approach for the GD effect, i.e., Eq.~(\ref{eq35b}) with $\kappa=$ constant and $\beta_1=1$, which according to the present line of thinking represents another extreme case. In this approach $S_{\lambda}$ varies in depth at each $\theta$ as predicted by the local $[T_{\rm eff}(\theta),\log g_{\rm eff}(\theta)]-$dependent plan-parallel models of stellar atmospheres. 
\end{itemize} 

   To account for the combined effects 1) and 2), we have calculated rotationally broadened \ion{He}{i}\,4471 and \ion{Mg}{ii}\,4481 line profiles with FASTROT \citep{frem05} adapted for the above listed atmospheric structures in model stars with $pnrc$ fundamental parameters $T_{\rm eff}=$ 12\,500 K and 22\,000~K ($pnrc=$ parent-non-rotating-counterpart), both having $\log g=$ 4.0, rotating at $\eta=0.99$, and seen at $i=30\degr$, 60\degr and 90\degr.  In FASTROT the sensitivity to the local formation conditions of spectral lines evoked in 1) are taken into account automatically. Although the 
\ion{He}{i}\,4471 and \ion{Mg}{ii}\,4481 lines always appear blended in stellar spectra, here we purposefully study these lines in isolation. In this way we can view these lines as representative of deviations that other isolated spectral lines with differentiated sensitivities to local formation conditions may carry on the fundamental parameter determination. Figure~\ref{fig_19} shows the line profiles of isolated \ion{He}{i}\,4471 and \ion{Mg}{ii}\,4481 lines according to the above assumptions a), b) and c).\par
   Apart from the obvious differences that can be seen in Fig.~\ref{fig_19} on the line profile predictions according to the type of GD adopted and the behavior of the line source function,  for the intermediate effective temperatures there are significant differences in the predictions based on assumptions a) and c) when inclination angles are $i\gtrsim60\degr$, under which most stars are seen. Accordingly, conflicting determinations of apparent fundamental parameters can result. \par 
   To sketch these conflicting determinations, we used the equivalent widths of the \ion{He}{i}\,4471 and \ion{Mg}{ii}\,4481 calculated in Fig.~\ref{fig_19} and interpolated the apparent $T^{\rm app}_{\rm eff}$ and $\log g^{\rm app}_{\rm eff}$ in the model relations between these lines and the effective temperature and gravity in non-rotating stars shown in \citet{frem05} (see their Fig.~4). The obtained results are shown in Table~\ref{tab_app}, which includes the corresponding $V\!\sin i$ determined using the classical Fourier transform method.\par 
   Several conclusions can be drawn from these results:\par
\begin{enumerate}
\item The apparent $T^{\rm app}_{\rm eff}$ and $\log g^{\rm app}_{\rm eff}$ rather strongly depend on the assumption made on the behavior of the radiation source function in GD atmospheres. As expected, the line profiles issued from model c) lead to the lower estimates of $T^{\rm app}_{\rm eff}$.   
\item The $V\!\sin i$ derived with the \ion{Mg}{ii}\,4481 line are systematically larger than those obtained with the \ion{He}{i}\,4471 line, which is a consequence of the differentiated sensitivity of these lines to the local formation 
conditions.
\item As expected, and already well known, the gravity-darkened models produce lower values of $V\!\sin i$ than models 
without gravity darkening.
\item Models with gravity darkening where the line source functions vary with $\theta$ produce the lowest $V\!\sin i$
parameters. 
\item The differences $\Delta V\!\sin i=$ $|V\!\sin i({\rm Mg})-V\!\sin i({\rm He})|$ are larger in those gravity-darkened models where the line source functions vary with $\theta$. 
\end{enumerate}

  Some of the assumptions made above on the physical structure of stellar atmosphere in rapidly rotating stars can unleash horizontal diffusion of light that has to be treated properly to determine the stable atmospheric thermal structure. Detailed two-dimensional radiation transfer calculations in rotationally deformed stars, similar 
to those started by \citet{pust70} and \citet{hadr92}, might be able to tackle these questions and predict reliable dependencies of $S_{\lambda}$ with $\theta$ as well as gravity darkening relations consistent with the right thermal structures of atmospheres. From these we may then expect to have reliable stellar fundamental parameters and line profiles required to make some progress in determining the properties of the stellar surface rotation.\par 

\begin{figure*}[]
\center\includegraphics[scale=0.75]{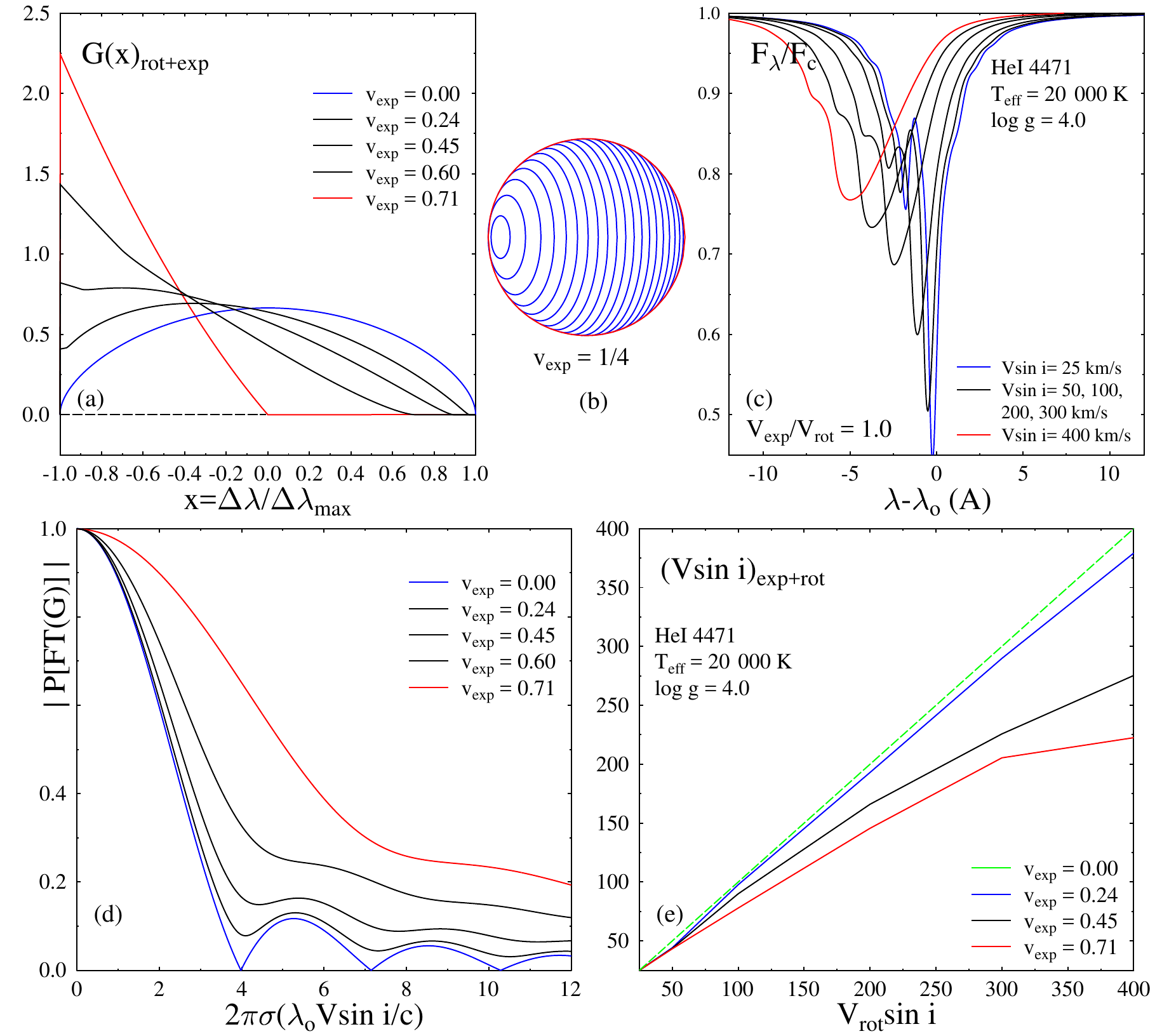}
\caption{\label{fig_21} (a) Tested truncated rotation-expansion broadening functions (REBF) for several velocity ratios $v_{\rm exp}=V_{\rm exp}/[V^2_{\rm exp}+V^2_{\rm rot}]^{1/2}]$. (b) Loci of points with the same projected velocity in the interval of Doppler displacements $-1\leq\Delta\lambda/\Delta\lambda_{\rm M}\leq V_{\rm rot}/[V^2_{\rm exp}+V^2_{\rm rot}]^{1/2}$ that contribute to the broadening of line profiles when $v_{\rm exp}=0.24$. (c) \ion{He}{i}\,4471 line profiles broadened for several input rotation velocities $V_{\rm rot}\!\sin i$ and velocity ratio $V_{\rm exp}/V_{\rm rot}=1.0$. (d) FT of the REBF. (e) Resulting $V\!\sin i_{\rm exp+rot}$ parameters produced in the frame of an expanding and rotating atmosphere against the model input $V_{\rm rot}\!\sin i$ for several ratios $v_{\rm exp}$.}
\end{figure*}

\subsection{Asymmetric rotational broadening function}\label{glob_rotbr}
   
  The lines due to a variety of elements and their different ionization states in a given analyzed wavelength range are formed in atmospheric layers characterized by different physical conditions and, in particular by velocity fields entertained by non-radial pulsations, convection movements, more or less expanding layers of nascent/starting winds, etc. When the rotational broadening function (RBF) is determined by deconvolution of all spectral lines present in that spectral range, there is a small chance that its shape is symmetric \citep{reyn1,reyn2}. It can be shown that, independent of the sign of the skewness of RBF, we always obtain underestimated $V\!\sin i$ parameters. To quantify this effect, we distorted the classic symmetric RBF of rigid rotators $G_{\rm S}(\Delta\lambda)$ using a linear function of the normalized wavelength displacement 
$\Delta\lambda/\lambda_{\rm M}$ as follows:

\begin{equation}
G_{\rm A}(\Delta\lambda/\lambda_{\rm M}) = G_{\rm S}(\Delta\lambda/\lambda_{\rm M})_{\rm s}\times[1+a\times(\Delta\lambda/\lambda_{\rm M})],
\label{eq38}
\end{equation}

\noindent where ${-1\leq a\leq1}$ is the free distortion-parameter and $\Delta\lambda_{\rm M}$ is the wavelength displacement induced by the equatorial rotation velocity. Figure~\ref{fig_20}a shows a set of such asymmetric RBF. Figure~\ref{fig_20}b shows the \ion{He}{i}\,4471 line ($T_{\rm eff}=$ 20\,000 K, $\log g=$ 4.0) rotationally broadened with Eq.~(\ref{eq38}) with ${a=-1}$ for several apparent equatorial rotational velocities $V\!\sin i$. The lines are not only broadened, but they become increasingly shifted the higher the rotational velocity. Figure~\ref{fig_20}c shows the FT of the asymmetric RBF $G_{\rm A}(\Delta\lambda)$, where the maxima of lobes are the same for all $V\!\sin i$, but the zeros of the symmetric RBF $G_{\rm S}(\Delta\lambda)$ are transformed into functional minima that are slightly displaced as soon as $a\neq0$. Figure~\ref{fig_20}d shows the relation between the $(V\!\sin i)_{\rm A}$ implied by the distorted spectral lines and the true $(V\!\sin i)_o$ value, where the underestimation of the projected rotational velocity appears as an increasing function of the distortion parameter $a$ and $(V\!\sin i)_o$.\par  
   We can then conclude that when a RBF is obtained by deconvolution of a rather wide spectral range, the supposedly more precise RBF introduces deviations that lead to underestimated parameters $V\!\sin i$. Generally, the common practice is to symmetrize the RBF to derive  $V\!\sin i$ from its TF. However, the symmetrization of a RBF changes its nature; this does not necessarily lead to the expected right $V\!\sin i$ value, as follows from the comparisons of $V\!\sin i$ values made in \citet{coll95}, where the authors adopt as examples the line broadening produced by elliptical and parabolic RBF functions.\par 

\begin{figure*}[] 
\center\includegraphics[scale=0.75]{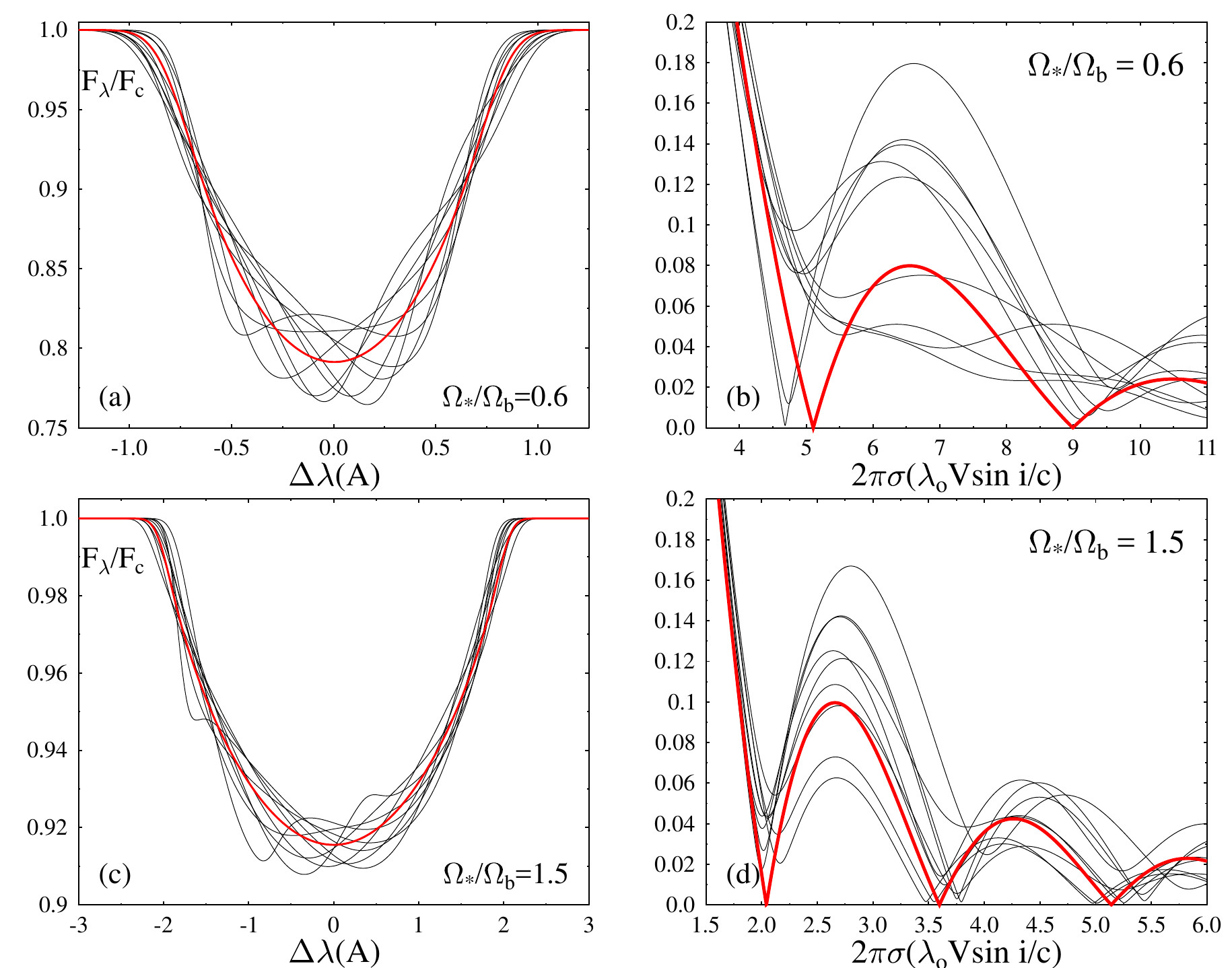} 
\caption{\label{fig_22} (a) Model line profiles of the primary star broadened by rotation and perturbed by tidal interactions in a binary system with sub-synchronous rotation ($\Omega_*$ stellar angular velocity; $\Omega_{\rm b}$ circular orbital angular velocity). (b) FT of spectral lines shown in (a). (c) Model line profiles of the primary star broadened by rotation and perturbed by tidal interactions in a binary system with super-synchronous rotation. (d) FT of spectral lines shown in (d). The red lines are either for the averaged line profiles, which represent the tidally unperturbed rotationally broadened line profile, or for the FT of the averaged line profile.}
\end{figure*}

\subsection{Effects carried by expanding layers}\label{exp_lay}

  During the seventies and eighties a large amount of literature dealt with the spectral line formation problem in moving stellar atmospheres \citep[see references in][]{sobo60,mih78,thom82,thom83,kalk84,mihmih84,kuz85,kalk87,thom88,
senw98,stee02,hubmih03,cann12,hubmih14}. These velocity fields were put into evidence mainly through spectroscopic observations carried out with satellites in the far-UV spectral range. The far-UV line asymmetries and the multicomponents of certain transitions were associated with mass ejections through winds, whose velocities are accelerated by the radiation pressure only once some layers start expanding. The presence of stellar winds thus implies the existence of expanding layers somewhere deeper in the atmosphere. In the literature, there are also claims of evidence of velocity fields detected in the visual spectral range of single B-type stars \citep{furyug}. The typical outbursts and fadings in Be stars, which are thought to be the consequence of huge discrete ejections of mass \citep{cook95,hb98,kell02,menn02}, also imply that there are periods with outward accelerated layers in the stellar atmospheres \citep{hb00,dw06}.\par  
  The combination of expanding velocities with rotation produces broadened and blueshifted spectral lines that can be schematically described with rotation-expansion broadening functions (REBF). These functions are asymmetric, and depending on the ratio of characteristic velocities $V_{\rm exp}/V_{\rm rot}$, they can become truncated. This phenomenon was studied analytically in the frame of uniform expansions by \citet{duvkar}. Detailed models of stellar atmospheres in rotation and expansion fields with velocity gradients can be found in \citet{mihls}.\par 
  We used the relationships obtained by \citet{duvkar} to infer the order of magnitude of effects produced 
on the $V\!\sin i$  determination by expanding velocity fields. Figure~\ref{fig_21}a shows a series of REBF obtained for different velocity ratios $v_{\rm exp}=V_{\rm exp}/[V^2_{\rm exp}+V^2_{\rm rot}]^{1/2}]\leq1$ ($V_{\rm rot}=V\!\sin i$ where  $i=\pi/2$). Thus, $v_{\rm exp}=1$ corresponds to pure expansion,  $v_{\rm exp}=0$ is for pure rotation, and $v_{\rm exp}=1/\sqrt{2}$ identifies $V_{\rm exp}/V_{\rm rot}=1$. Figure~\ref{fig_21}b depicts the loci of points of isoradial velocity contours for $v_{\rm exp}=1/4$, where the associated Doppler displacements belong to the interval $-1\leq\Delta\lambda/\Delta\lambda_{\rm M}\leq V_{\rm rot}/[V^2_{\rm exp}+V^2_{\rm rot}]^{1/2}$. Figure~\ref{fig_21}c shows \ion{He}{i}\,4471 model line profiles ($T_{\rm eff}=$ 20\,000 K, $\log g=$ 4.0) broadened with several values of $V_{\rm rot}$ and $V_{\rm exp}/V_{\rm rot}=1$, where the blueshift produced by the expansion velocity component is clearly apparent. In Fig.~\ref{fig_21}d are shown the particular shapes of the FTs of \ion{He}{i}\,4471 lines broadened and deformed according to the indicated velocity ratios. The relations between the measured $V\!\sin i$ parameters, using the blueshifted and rationally broadened \ion{He}{i}\,4471 lines against the true $V_{\rm rot}\!\sin i$ for several ratios $v_{\rm exp}$, are shown in Fig.~\ref{fig_21}e.\par 
   The results shown in Fig.~\ref{fig_21} suggest at least two conclusions: 1) the underestimation of $V\!\sin i$ can be quite significant if the studied Be stars have been observed at phases when their atmospheres were driven by expanding velocity fields; 2) the blueshifts of lines can be more or less chaotic; they could partially explain irregular radial velocity drifts, which sometimes are suspected due to undisclosed binaries. Unfortunately, there is not enough systematic information to establish a distribution function of expanding velocities in atmospheres of Be stars to estimate their effect on the distribution $\Phi(u)$ and put to test the function $\phi(\alpha)$ [or $\phi(\delta)$] studied in Sect.~\ref{sect_23}.\par 
   In the present discussion we have assumed that the entire atmosphere of stars undergoes uniform expansion. Nonetheless, this movement may possibly concern only an equatorial strip, as suggested by the correlation against the $V\!\sin i$ of light outbursts and fadings in Be stars obtained by \citet{hb98}. In such a case the underestimation of the $V\!\sin i$ parameter can be smaller than predicted here. \par  
 
\subsection{Tidal interactions in binary systems}\label{bin_tidal}  

   A large percentage  of massive and intermediate mass stars are in binary systems. Two interactions in binary systems can then perturb the determination of the $V\!\sin i$ of individual components: mutual irradiation and tidal interaction.\par 
   Owing to the mutual irradiation and radiation pressure that become significant in close binary systems \citep{cranm93,drech95,phil02,palate13}, line spectra can change somewhat as a function of the orbital phase and inclination angle of the system. In this discussion we omit the mutual irradiation and radiation pressure.\par      
   \citet{more1,more2} have studied the variation of photospheric spectral line profiles induced by the tidal deformations in binary systems, which can also induce differential rotation. Their results show that there are traveling bumps produced mainly by the azimuthal components of the velocity perturbations, which are superimposed on the rotationally broadened profiles. The number and strength of these bumps depend on the orbital phase and they are particularly strong in non-synchronous systems.  Dr. G. Koenisberger has graciously given us the entire series of  calculations of line profiles published in \citet{more2}, which we have shifted to a common central rest wavelength and applied the FT method to determine the apparent rotational velocity $V\!\sin i$ for each of them. Figure~\ref{fig_22}a shows the centered rotationally broadened and tidally perturbed line profiles in the case of a sub-synchronous rotation with $\Omega_*/\Omega_{\rm b}=0.6$ ($\Omega_*$ is the stellar angular velocity; $\Omega_{\rm b}$ is the circular orbital angular velocity) and where the primary star has $V\!\sin i=$ 130 km~s$^{-1}$.  The respective FTs of these lines are shown in Fig. 17b. Figures~\ref{fig_22}c,d are equivalent to Figs.~\ref{fig_22}a,b, but depict the case of a super-synchronous rotation with $\Omega_*/\Omega_{\rm b}=1.5$. In these figures, the red lines represent the averaged line profiles, which very closely correspond to the unperturbed rotationally broadened line, and their respective FT. In Table~\ref{tab_5} are reproduced the $V\!\sin i$ parameters obtained from each individual line profile, which enables us to appreciate the apparent phase-dependent change of $V\!\sin i$. \par 

\begin{table}[]
\centering
\caption[]{\label{tab_5} $V\!\sin i$ parameters in km/s determined using line profiles perturbed by tidal interactions in binary systems.}
\tabcolsep 2.0pt
\begin{tabular}{c|c|cccccccccc}
\hline
 $\Omega_*/\Omega_{\rm b}$ & $(V\!\sin i)_o$ & \multicolumn{9}{c}{perturbed $V\!\sin i$} & \\
\hline
 0.6 & 130 & 142 & 138 &  80 & 120 & 134 & 140 & 137 & 121 &  94 & 132 \\ 1.5 & 130 & 130 & 132 & 133 & 125 & 123 & 130 & 134 & 133 & 126 & 129 \\\hline
\end{tabular}
\end{table}

\subsection{Presence of circumstellar envelopes or discs}\label{env_disc} One of the most outstanding hallmarks of Be stars is the presence of circumstellar envelopes or discs (CD), where their characteristic emissions and shell absorptions are raised. A large amount of literature describes the formation of these spectral features \citep[cf.][]{thom82,marta13}. The $V\!\sin i$ determination can then be affected by variable emissions and/or absorptions superimposed on the lines used to determine this parameter. These effects on the 
\ion{He}{i}\,4471 line were discussed in \citet{ball95}, where the detected emission can carry overestimations on the order of $30\pm20$~km~s$^{-1}$ in Be stars with effective temperatures $15\,000\lesssim T_{\rm eff}\lesssim28\,000$~K.
The \ion{Mg}{ii}\,4481 line, which is also of frequent use to measure the stellar rotation, can be affected either by circumstellar emission or shell absorption. The latter deepens the line absorption leading to underestimated values of $V\!\sin i$. In the visible spectral region of Be stars, no line can be considered entirely free from circumstellar effects, but depending on the sub-spectral type some lines, such as
\ion{Si}{i} and \ion{Si}{ii} lines, are apparently less damaged.  However, when some lines seem to be good photospheric witnesses because of the gravity darkening effect their formation can be favored either on polar or equatorial regions as shown in Sect.~\ref{rad_sf}. Only a careful modeling of such transitions can then lead to more reliable estimates of the  $V\!\sin i$ parameters in Be stars.\par
 Be stars are the epitome of stellar rapid rotators. However, their spectra can be marred by numerous disturbances produced by circumstellar envelopes or discs, which make their interpretation difficult and uncertain. These interpretations concern all insights we can draw from spectra on the surface and/or internal distribution of the angular velocity. As much it may concern the study of the angular momentum 
distribution in rapidly rotating stars, we should not neglect Bn stars because they are very rapid rotators and do not display line emission components and/or shell absorptions in their spectra. Moreover, there are statistical insights suggesting that these stars may represent a precursor stage to the  Be phase of late B-type stars \citep{zor_imf,zor05}. Finally, most Bn stars are of late B spectral type, which are more numerous and closer to us than early-type Be stars.\par 

\section{Comments and conclusions}\label{comm_concl}  

   Many effects that obliterate the actual information carried by the $V\!\sin i$ and/or $V$ parameters on the rotation cannot be studied individually for each Be star. We have then considered them statistically by studying the distribution of ratios $V\!\sin i/V_{\rm c}$ and $V/V_{\rm c}$.\par
   In Paper I \citep{zor16a} we studied a sample of 233 Galactic classical Be stars and obtained the distribution of their $V/V_{\rm c}$ by considering that the inclination angles $i$ are distributed at random. We redetermined the fundamental parameters of the studied objects to correct the $V\!\sin i$ from its underestimation due to the gravity darkening effect. We also considered the overestimation of $V\!\sin i$ due to macroturbulent 
velocities.\par 
   In the first part of the present contribution (Sect.~\ref{dif_rot}), we assumed that the atmospheres of Be stars undergo differential rotation. Sects.~\ref{why_difrot} and \ref{rot_broad} describe some properties of lines broadened by surface rotations laws characterized by angular velocities accelerated either toward the pole or the equator. We have shown that the main characteristics of the line profiles broadened by these laws can be reasonably accounted for with  the Maunder relation, which depends on a single free parameter $\alpha$. \par  
  In Sect.~\ref{eff_vsini} we studied the effect produced by the differential rotation on the value of the 
$V\!\sin i$ parameter and concluded that rotation laws in which the angular velocity is accelerated from the pole toward the equator ($\alpha<0$) tend to produce $V\!\sin i<V_{\rm eq}\!\sin i$, where $V_{\rm eq}$ is the li\-ne\-ar rotational velocity of the equator, while laws with accelerated from the equator toward the pole ($\alpha>0$) lead to $V\!\sin i>V_{\rm eq}\!\sin i$. Moreover, we have shown that me\-thods like the Fourier transform produce unreliable values of $V\!\sin i$ because the zeroes of a putative rotationally broadening function cannot be associated with those inherent to a broadening function derived for rigid rotators. We noted that for stars with differential rotation the line broadening function cannot be defined, since neither the rotation law nor the inclination of the star are known.\par  
  To complete the discussion on the effects carried by the differential rotation, in Sect.~\ref{eff_phiu} we estimated the effect it can produce on the distribution of ratios $V/V_{\rm c}$ of true rotational velocities. Owing to the displacement produced by the differential rotation of the distributions of apparent velocity ratios with respect to the distribution of the actual equatorial velocity ratios, we concluded that if a dominant number of Be stars rotate with laws characterized by $\alpha<0$, the number of rotators with $V_{\rm eq}\gtrsim0.9V_{\rm c}$ will be larger than expected from the observed distribution $\Phi(V/V_{\rm c})$. However, they could be lower if the Be stars had on average $\alpha>0$. \par      
  In the second part of the present contribution (Sect.~\ref{uncert_vsini}) we examined a number of conceptual and measurement uncertainties other than differential rotation that affect the determination of the $V\!\sin i$ parameter, which are not currently evoked in the studies of stellar rotation. These include \par 

\begin{itemize}
 
\item[$i)$] The angular momentum content, which can introduce increased geometrical deformations and subsequent gravity darkening contrasts; 
\item[$ii)$] Bi-valued relations between the line broadening and $V\!\sin i$, which can be due to surface rotation laws accelerated from the equator toward the pole; 
\item[$iii)$] The gravity darkening effect, which cannot be reliably characterized with a constant gravity darkening exponent as it is the current practice, and we question whether the radiation emitted by a rotationally deformed star can be reliably accounted for with classical models of stellar atmospheres responding to the local effective temperatures and gravities; 
\item[$iv)$] Asymmetric rotational broadening functions, which are determined by deconvolution of wide spectral ranges;\item[$v)$] Effects caused on spectral lines by expansion velocities in stellar atmospheres, which can lead to underestimated $V\!\sin i;$
\item[$vi)$] Deformations of line profiles by tidal interactions in binary systems, which introduce changes of $V\!\sin i$ correlated with the orbital phase; 
\item[$vii)$] Overestimations or underestimations of the $V\!\sin i$ due to the presence of circumstellar envelopes or discs.
 
\end{itemize}
 
  We note that from the numerical values of errors affecting the estimates of the $V\!\sin i$ of Be stars due to phenomena $v)$ and $vii)$, we could expect they can partially compensate each other as they act in opposite senses.\par 
  Although there is not enough data to estimate the effects enumerated in Sect.~\ref{uncert_vsini} to put to test the probability distribution $\phi(\alpha),$ progress regarding the $V\!\sin i$ determination can be made as follows: a) by studying highly resolved spectra over large wavelength intervals with spectral lines that are good tracers of formation conditions (temperatures, densities, velocity fields); b) by obtaining models of the observed spectral region with two-dimensional radiation transfer calculations in rotationally deformed atmospheres, including differential rotation, which solely can predict consistent gravity darkening laws; and c) by observing Bn stars to avoid spectral perturbations due to circumstellar envelopes and possibly huge expansion velocity fields in the atmospheres.\par
  Thanks to the super-resolution capabilities of modern spectrointerferometry, combined with high resolution spectroscopy, it could be possible to obtain direct information on the inclination angle $i$ and the differential parameter $\alpha$ \citep{domic04,vinic06,zor11,omar13,souz14}.\par    The knowledge of the degree of surface differential rotation and taking the effects mentioned in the list into account above will help, in particular, to decide whether or not the Be phenomenon mostly rely on the critical 
rotation.\par     
   Since differential rotation can be present in rapidly rotating objects other than Be stars, increased attention should be put to the observational aspects of this phenomenon. In fact, every observational indication on this phenomenon can be critical to deepen our understanding of the mechanisms involved in the angular momentum redistribution in stars and 
of the concomitant mixing phenomena of chemical elements, which help to predict the frequency of several stellar populations currently used to scrutinize the evolution of galaxies and/or the massive star formation rates \citep{maemey95,maemey04}.\par
    
\begin{acknowledgements} 
We thank Dr. Gloria Koenigsberger for having provided us with line profiles perturbed by tidal interaction in binary systems. We are thankful to the referee for his(her) useful and constructive criticisms and suggestions that significantly helped to correct and put in better focus some subjects discussed in this work. We are also indebted to
Amy Mednik for the rapid and efficient language editing of this paper.  
\end{acknowledgements}

\bibliographystyle{aa} 
\bibliography{28761}    

\begin{appendix}
\section{Binarity}\label{nobin}   

 The total radial velocity of a given point in the surface of a stellar component in a binary system has two components: one is due to the stellar rotation proper and the other to the orbital motion. 
In Paper I the second component was supposed to induce an additional broadening of lines through a differential Doppler shift. Actually, two points in the stellar surface situated at the same latitude from the apparent pole in an axially symmetric object and  symmetrically, at both sides of the meridian plane that contains the rotation axis and the line of sight, have the same radial velocity in the direction toward the observer. Thus, the suggested differential Doppler shift does not exist and the effect due to binarity discussed in Paper I on $V\!\sin i$ has to be dismissed (J.Z.). \par 
\end{appendix}

\end{document}